\documentclass{ws-mplb}
\usepackage{epsfig}

\def\avg#1{\langle#1\rangle}

\def\be{\begin{equation}}       \def\ee{\end{equation}}
\def\bea{\begin{eqnarray}}      \def\eea{\end{eqnarray}}

\def\nn{\nonumber} 
\def\pp{\parallel}

\begin{document}

\markboth{Congjun Wu}{Unconventional  Bose-Einstein Condensations
Beyond the ``No-node'' Theorem}

\catchline{}{}{}{}{}
%
%%%%%%%%%%%%%%%%%%%%%%%%%%%%%%%%%%%%%%%%%%%%%%%%%%%%%%%%%%%%%%%%%%%%

\title{Unconventional Bose-Einstein Condensations 
Beyond the ``No-node'' Theorem
%\footnote{For the title, try not to 
%use more than 3 lines. Typeset the title in 10 pt 
%Times roman, uppercase and boldface.}
}

\author{\footnotesize Congjun Wu
%\footnote{Typeset names in 
%10~pt Times roman, uppercase. Use the footnote to indicate 
%the present or permanent address of the author.}
}

\address{ Department of Physics, University 
of California, San Diego \\  La Jolla, CA 92093-0319, USA. \\
wucj@physics.ucsd.edu}

\maketitle

\begin{history}
\received{(Day Month Year)}
\revised{(Day Month Year)}
\end{history}

\begin{abstract}
Feynman's ``no-node'' theorem  states that the conventional 
many-body ground-state wavefunctions of bosons in the coordinate
representation is positive-definite.
This implies that time-reversal symmetry cannot be spontaneously broken.
In this article, we review our progress in studying a class of
new states of unconventional Bose-Einstein condensations 
beyond this paradigm.
These states can either be the long-lived metal-stable states of ultra-cold 
bosons in high orbital bands in optical lattices as a result of
the ``orbital-Hund's rule'' interaction, or the ground states of
spinful bosons with spin-orbit coupling linearly dependent on momentum.
In both cases, Feynman's argument does not apply.
The resultant many-body wavefunctions are complex-valued and thus
break time-reversal symmetry spontaneously. 
Exotic phenomena in these states include the Bose-Einstein condensation 
at non-zero momentum, the ordering of orbital angular momentum moments, 
the half-quantum vortex, and the spin texture of skyrmions.
\end{abstract}

\section{Introduction}
\label{sect:intro}

In Feynman's statistical mechanics textbook, it is stated that the many-body
ground state wavefunctions of bosons are {\it positive-definite}
in the coordinate representation provided no external rotation is applied 
and interactions are short-ranged \cite{feynman1972}.
The proof is very intuitive: due to the time-reversal (TR) symmetry, 
the ground state wavefunction $\Psi(\vec r_1, ...., \vec r_n)$ 
can be chosen as real.
If it is not positive-definite, {\it i.e.}, it has nodes, the following 
surgery can be done to lower its energy.
We first take its absolute value of $|\Psi(\vec r_1, ...., 
\vec r_n)|$, whose energy expectation value is exactly the
same as that of $\Psi(\vec r_1, ...., \vec r_n)$.
However, such a wavefunction has kinks at node points.
Further smoothing the kinks  results in a positive-definite
wavefunction, and the kinetic energy is lowered by softening the 
gradients of the wavefunction.
Although the single body potential energy and the two-body interaction 
energy increase, they are small costs of a high order compared
to the gain of the kinetic energy.
We can further conclude that the ground state is non-degenerate
because two degenerate positive-definite wavefunctions cannot be 
orthogonal to each other.

This ``no-node'' theory is a very general statement, which
applies to almost all the well-known
ground states of bosons, including the superfluid, Mott-insulating, 
density-wave, and even super-solid ground states.
It is also a very strong statement, which reduces the generally 
speaking complex-valued many-body wavefunctions 
to be positive-definite distributions.
This is why the ground state properties of bosonic systems, such as $^4$He,
can in principle be exactly simulated by the quantum Monte-Carlo method
free of the sign problem.
Furthermore, this statement implies that the ordinary ground 
states of bosons, including Bose-Einstein condensations (BEC) and 
Mott-insulating states, cannot spontaneously break 
time-reversal (TR) symmetry, since TR transformation for the single
component bosons is simply the operation of the complex conjugation.

It would be exciting to search for exotic emergent states of bosons 
beyond this ``no-node'' paradigm, whose many-body wavefunctions
can be complex-valued with spontaneous TR symmetry breaking.
Since properties of complex-valued functions are much richer than
those of real-valued ones, we expect such states can exhibit more intriguing 
properties than the ordinary ground states of bosons.
For this purpose, we have recently made much progress with two
different ways to bypass Feynman's argument, including the meta-stable states
of bosons in the high orbital bands of optical lattices 
\cite{liu2006,wu2006,stojanovic2008}
and multi-component bosons with spin-orbit coupling linearly 
dependent on momentum \cite{wu2008c}.

Clearly, the ``no-node'' theorem is a ground state property which does not
apply to the excited states of bosons.
The recent rapid development of optical lattices with ultra-cold
bosons provides a wonderful opportunity to investigate the
meta-stable states of bosons pumped into high orbital bands.
Due to the lack of dissipation channels, the life time can be
long enough to develop inter-site coherence \cite{isacsson2005,mueller2007}.
We have shown that the interaction among orbital bosons are characterized 
by the ``{\it orbital Hund's rule}'' \cite{liu2006,wu2006}, which
gives rise to a class of complex superfluid states by developing
the on-site orbital angular momentum (OAM) moments.
In the lattice systems, the inter-site hoppings of bosons lock the 
OAM polarization into regular patterns depending the concrete 
lattice structures.

Furthermore, the ``no-node'' theorem even does not apply to the ground
states of bosons if their Hamiltonians linearly depend on momentum,
i.e., the gradient operator.
A trivial example is the formation of vortices in Bose condensates
with the external rotation in which the Coriolis force is represented 
as the vector potential linearly coupled to momentum.
However, in this case TR symmetry is explicitly broken.
A non-trivial example is that bosons with spin-orbit (SO) coupling, whose 
Hamiltonians also linearly depend on momentum and are TR invariant.
The invalidity of the ``no-node'' theorem can also give rise to
complex-valued ground state wavefunctions.

Although $^4$He is spinless and most bosonic alkali atoms are too 
heavy to exhibit the relativistic SO coupling in their center-of-mass 
motion, SO coupling can be important in exciton systems 
in two dimensional quantum wells.
We have shown that the Rashba SO coupling in the conduction electron
bands induces the same type of SO coupling in the center-of-mass motion 
of excitons.
In a harmonic trap, the ground state condensate wavefunction
can spontaneously develop half-quantum vortex structure and
the skyrmion type of the spin texture configuration
\cite{wu2008c}.
On the other hand, effective SO coupling in boson systems can be
induced by laser beams, which has been investigated in 
several publications by other groups in 
Refs \cite{lin2008,juzeliunas2008,stanescu2007,stanescu2008}.

In the following, we will review our work in both directions
outlined above including many new results unpublished before.
In Section \ref{sect:boson}, we explain the characteristic feature
of interacting bosons in high orbital bands, the ``orbital Hund's rule'',
and the consequential complex-superfluid states with the ordering
of on-site OAM moments.
The ordering of OAM moments in the Mott-insulating
states is also investigated.
In Section \ref{sect:spinorbit}, we review the TR symmetry breaking
states of spinful bosons with spin-orbit coupling.
Interesting properties including 
half-quantum vortex, and the skyrmion-like spin-textures are studied.
Conclusions  are made in Section \ref{sect:conclusion}.

Due to the limit of space, we will not cover interesting related topics 
of orbital bosons, such as the nematic superfluid state \cite{isacsson2005} 
and the algebraic superfluid state \cite{xu2007},
and the orbital physics with cold fermions
\cite{wu2007,wu2008,wu2008a,wu2008b,zhang2008,umucalilar2008,wuzhai2008,zhao2008}, which has also aroused much research attention recently.

%*******************************************************************
%******************************************************************
%******************************************************************
\section{``Complex-condensation'' of bosons in high orbital bands
in optical lattices}
\label{sect:boson}
In this section, we will review the ``complex-condensation'' with TR
symmetry breaking, which is a new state of the $p$-orbital bosons.
This is an example of novel orbital physics in optical lattice with
cold bosons.
Below let us give a brief general introduction to 
orbital physics for the general audience.

Orbital is a degree of freedom independent of charge and spin. 
It plays important roles in magnetism, superconductivity, and 
transport in transition metal oxides \cite{imada1998,tokura2000,khaliullin2005}.
The key features of orbital physics are orbital degeneracy and
spatial anisotropy.
Optical lattices bring new features to orbital physics
which are not easily accessible in solid state orbital systems.
First, optical lattices are rigid lattices and free from the Jahn-Teller 
distortion, thus orbital degeneracy is robust.
Second, the meta-stable bosons pumped into high orbital bands 
exhibit novel superfluidity beyond Feynman's ``no-node'' theory
\cite{isacsson2005,liu2006,kuklov2006,wu2006,xu2007,xu2007a,stojanovic2008}. 
Third, $p$-orbitals have stronger spatial anisotropy 
than that of $d$ and $f$-orbitals, while correlation effects in 
$p$-orbital solid state systems (e.g. semiconductors) are not 
that strong.
In contrast, interaction strength in optical lattices is tunable.
We can integrate strong correlation with spatial anisotropy 
more closely than ever in $p$-orbital optical lattice systems
\cite{wu2007,wu2008,wu2008a,wu2008b,zhang2008}.

Recently, orbital physics with cold atoms has been attracting a great deal 
of attention\cite{scarola2005,isacsson2005,liu2006,kuklov2006,wu2006,xu2007,xu2007a,wu2007,alon2005,larson2008,browaeys2005,kohl2005,sebby-strabley2006,mueller2007}.
For orbital bosons, a series of theory works have been done 
\cite{isacsson2005,scarola2005,liu2006,kuklov2006,wu2006,xu2007,xu2007a,stojanovic2008,alon2005}, including  the illustration of
the ferro-orbital nature of interactions \cite{liu2006}, 
orbital superfluidity with spontaneous time-reversal symmetry breaking
\cite{liu2006,wu2006,kuklov2006,stojanovic2008}, the nematic
superfluidity\cite{isacsson2005,xu2007}.
The theory work on the $p$-orbital fermions is also exciting
including the flat band and associated strong correlation physics
in the honeycomb lattice \cite{wu2007,wu2008a,zhang2008},
orbital exchange and frustration \cite{wu2008b,zhao2008},
and topological insulators \cite{wu2008,umucalilar2008}. 

The experimental progress has been truly exciting 
\cite{browaeys2005,kohl2005,sebby-strabley2006,mueller2007}.
Mueller {\it et al.} \cite{mueller2007} have 
realized the meta-stable $p$-orbital boson systems by using the 
stimulated Raman transition.
The spatially anisotropic phase coherence pattern has been observed
in the time of flight experiments.
Sebby-strabley {\it et al.} \cite{sebby-strabley2006}
have successfully pumped bosons into excited bands in the
double-well lattice.
In addition, $p$-orbital Bose-Einstein condensation (BEC) has also
been observed in quasi one-dimensional exciton-polariton
lattice systems \cite{lai2007}.

Below we illustrate the important feature of interactions
between orbital bosons, the ``orbital Hund's rule'', which 
results in TR symmetry breaking.

%-------------------------------------------------------------
%------------------------------------------------------------
\subsection{Orbital Hund's rule of interacting orbital bosons}

\begin{figure}
\centering\epsfig{file=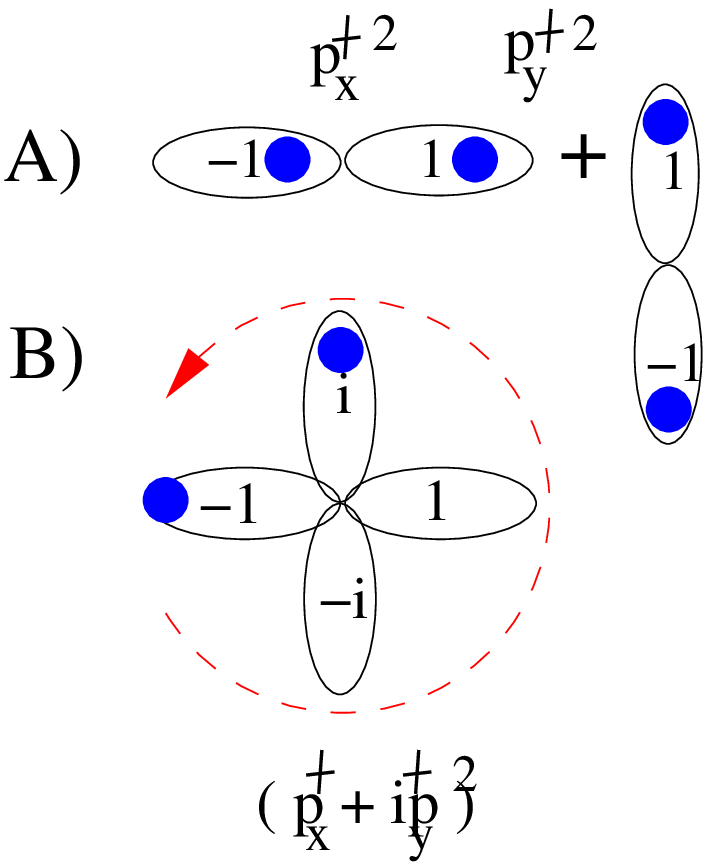,clip=1,width=0.21\linewidth,angle=0}
\hspace{3mm}
\centering\epsfig{file=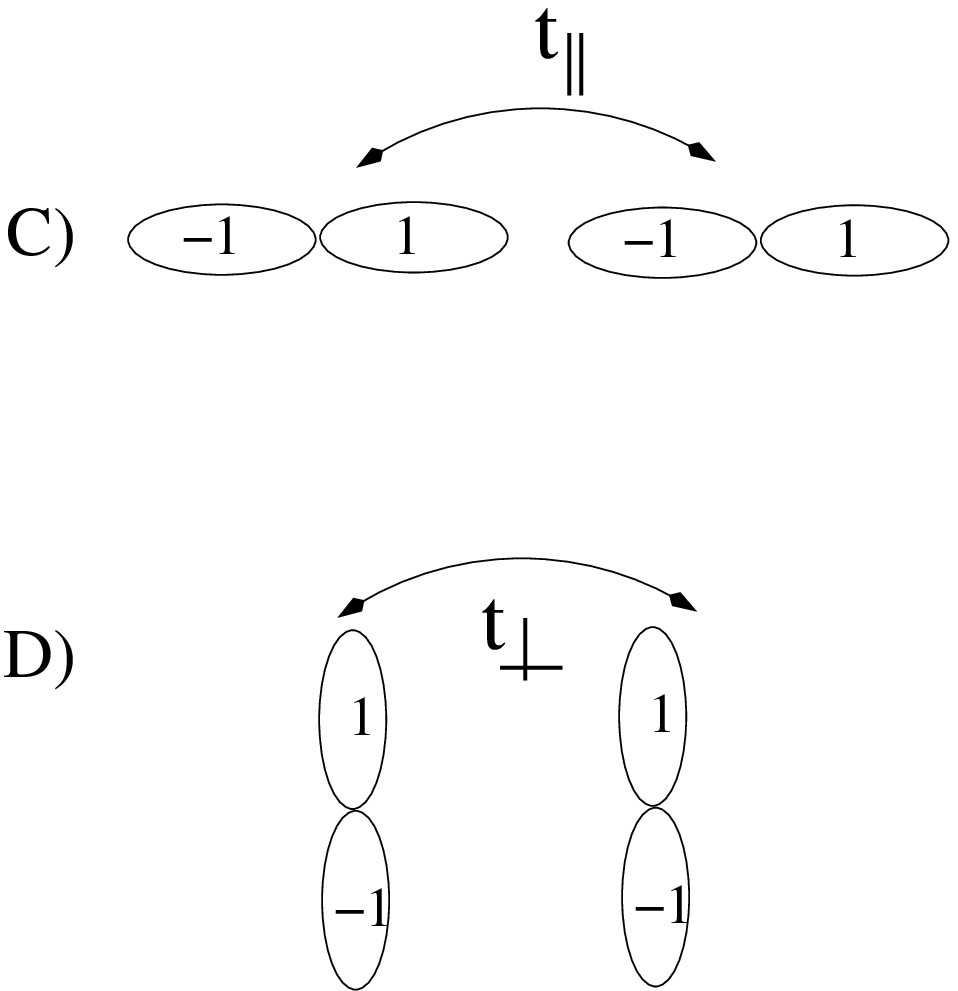,clip=1,width=0.25\linewidth,angle=0}
\caption{A single site problem with two spinless bosons occupying
$p_{x,y}$-orbitals: (A) the OAM singlet and (B) one of the OAM doublets. 
The latter is energetically more favorable than the former as a result of
the ``orbital Hund's rule''.
The bonding pattern of $p$-orbitals: (C) the $\sigma$-bonding 
and (D) the $\pi$-bonding.   
}
\label{fig:pboson}
\end{figure}

The most remarkable feature of interacting bosons in high orbital bands 
is that they favor to maximize their onsite orbital angular momentum (OAM) 
as explained below \cite{liu2006,wu2006}.
 
Let us illustrate this Hund's rule type physics through the simplest example: 
a single site problem with two degenerate $p_{x,y}$-orbitals 
filled with two spinless bosons.
Because bosons are indistinguishable, the Hilbert space for the two-body
states only contains three states, which can be classified 
according to their OAM: an OAM singlet as $\frac{1}{2}
(p_x^\dagger p_x^\dagger + p_y^\dagger p_y^\dagger) |0\rangle$ depicted in Fig 
\ref{fig:pboson}. A, 
and a pair of OAM doublets $ \frac{1}{2 \sqrt 2}(p^\dagger_x \pm i p^\dagger_y)^2 
|0\rangle$ with $L_z=\pm 2$ as depicted in Fig \ref{fig:pboson} B. 
Assuming a contact interaction $V=g\delta (r_1-r_2)$, the interaction 
energy of the former is calculated as $\frac{4}{3}U$ while that of 
the latter is $\frac{2}{3} U$ with the definition of
\bea
U=g\int d r^2 |\psi_{p_{x}} (r)|^4= g\int d r^2 |\psi_{p_{y}} (r)|^4.
\eea
In the OAM singlet state, bosons occupy polar (real) orbitals 
(e.g. $p_x, p_y$), whose angular distribution in real space is narrower
than that of the axial (complex) orbitals (e.g. $p_x \pm i p_y$ ) 
for the OAM doublets. 
By occupying the same axial orbital and therefore maximizing OAM, two 
bosons enjoy more room to avoid each other.

This ``ferro-orbital'' interaction is captured by the following multi-band
Hubbard Hamiltonian for the $p$-orbital bosons as
\bea
H_{int}=\frac{U}{2}\sum_{\vec r} \big\{n^2_{\vec r}-\frac{1}{3} 
 L_{z}^2 \big\},
\label{eq:bshamint}
\eea
where  $n=p^\dagger_x p_x+ p^\dagger_y p_y $ and
$L_z=-i(p^\dagger_x p_y- p^\dagger_y p_x)$.
The first term is the ordinary Hubbard interaction and the second term
arises because of the orbital degree of freedom.
In three dimensional systems in which all of the three $p$-orbitals
are present, we only need to replace $L_z$ with 
$\vec L^2=L_x^2+L_y^2+L_z^2$
and $L_{x,y}$ defined as $L_x=-i(p_y^\dagger p_z -p_z^\dagger p_y)$,
$L_y=-i (p_z^\dagger p_x -p_x^\dagger p_z)$.

When more than two bosons occupy a single site,  bosons prefer to 
go to the same single particle state by their statistical properties.
Again going into the same axial state and thus maximizing OAM 
can reduce their repulsive interaction energy.
This is an analogy to the Hund's rule of electron's filling in atomic
shells.
The first Hund's rule of electrons maximizes electron total spin to 
antisymmetrize their wavefunction, and the second Hund's rule further 
maximize their OAM to extend the spatial volume of wavefunction.
The key feature is that electrons want to avoid each other as possible
as they can.
For the orbital physics of spinless bosons, the same spirit applies 
with the maximizing of OAM.

Now we have understood the single site physics in which bosons
develop rotation.
From the symmetry point of view, it is similar to the $p_x+ip_y$
superconductors.

%--------------------------------------------------------
%---------------------------------------------------------
\subsection{Band structures of the $p$-orbital systems}
\label{subsect:band}

\begin{figure}
\centering\epsfig{file=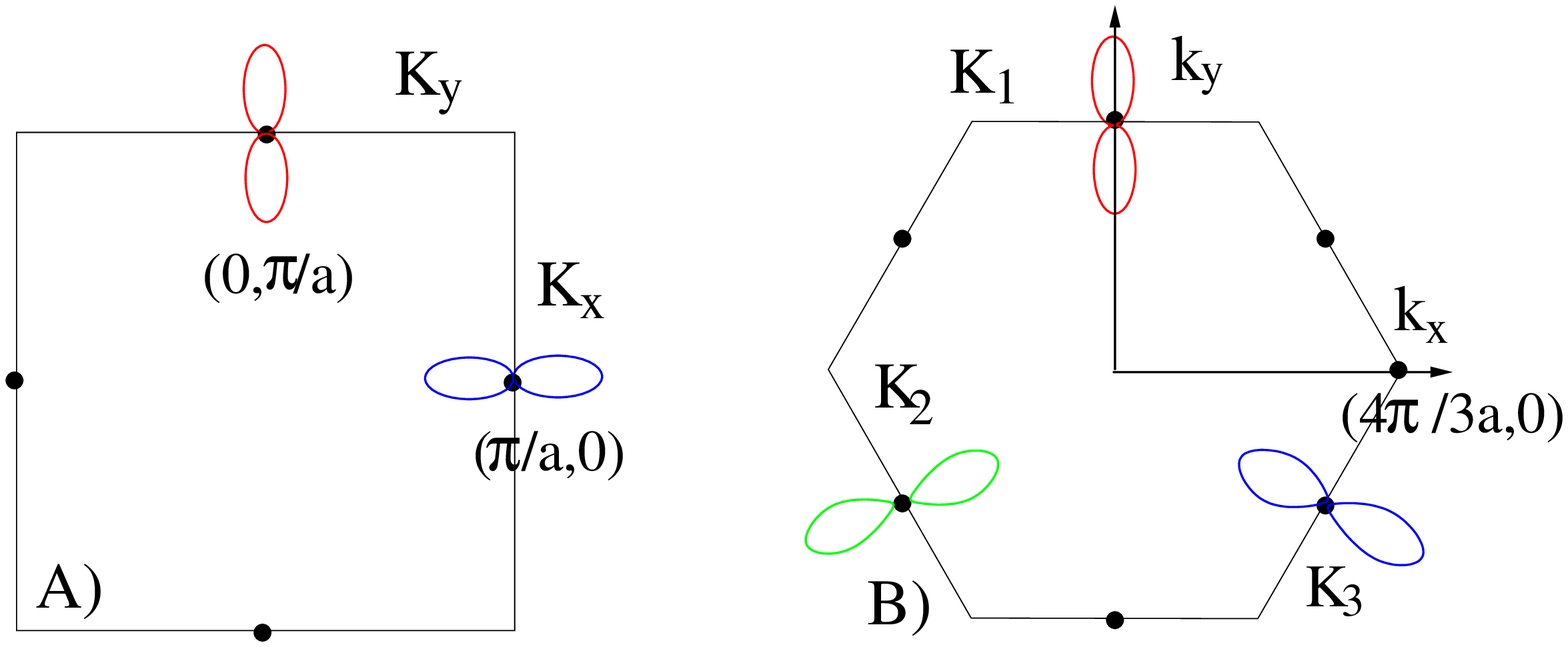,clip=1,width=0.6\linewidth,angle=0}
\caption{The $p$-orbital band structure in the two dimensions:
A) square and B) triangular  lattices.
In the square lattice, the band minima are located at
$K_x=(\pi/a_0,0)$ for the $p_x$-orbital band, and $K_y=(0,\pi/a_0)$
for the $p_y$-orbital band, respectively.
In the triangular lattice, the band minima are at three middle
points of Brillouin zone edges as $K_1=(0,\frac{2\pi}{\sqrt 3 a})$
and $K_{2,3}=(\frac{\pi}{a}, \pm\frac{\pi}{\sqrt 3 a})$.
The corresponding orbital configurations at these three band
minimal are polar-like and parallel to the momentum directions 
of $K_{1,2,3}$ respectively as depicted. 
}
\label{fig:band}
\end{figure}

The $p$-orbital Hamiltonian within the tight-binding approximation
can be written as
\bea
H_0= t_\parallel \sum_{\langle i j \rangle} [p^\dagger_{i, \hat e_{ij}} 
p_{j,\hat e_{ij}}+h.c.]-t_\perp \sum_{\langle i j \rangle}
[ p^\dagger_{i,\hat f_{ij}} p_{j,\hat f_{ij}} +h.c.],
\label{eq:bandsquare}
\eea
where the unit vector $\hat e_{ij}$ is along the bond orientation
between two neighboring sites $i$ and $j$ and $\hat f_{ij}=\hat z \times
\hat e_{ij}$ is perpendicular to $\hat e_{ij}$.
$p_{\hat e_{ij}}$ and $p_{\hat f_{ij}}$ are the projections of $p$-orbitals
along (perpendicular to) the bond direction respectively as defined
below
\bea
p_{\hat e_{ij}}= (p_x \hat e_x + p_y \hat e_y) \cdot \hat e_{ij}, \ \ \
p_{\hat f_{ij}}= (p_x \hat e_x + p_y \hat e_y) \cdot \hat f_{ij}.
\eea
The $\sigma$-bonding $t_\parallel$ and 
the $\pi$-bonding $t_\perp$ describe the hoppings along and perpendicular 
to the bond direction as depicted in Fig. \ref{fig:pboson} C and D, 
respectively.
The opposite signs of the $\sigma$ and $\pi$-bondings are due to the
odd parity of $p$-orbitals.
$t_\perp$ is usually much smaller than $t_\parallel$ in strong periodical 
potentials.

Eq. \ref{eq:bandsquare} exhibits degenerate band minima in 
both the square and triangular lattice\cite{liu2006,wu2006,wu2007}.
In the square lattice, the Brillouin zone (BZ) is a square with the edge 
length of $\frac{\pi}{a_0}$ where $a_0$ is the length between
the nearest neighbors.
The spectra read $\epsilon_{p_x}= t_\pp \cos k_x - t_\perp \cos k_y$
and $\epsilon_{p_y}=-t_\perp \cos k_x +t_\pp \cos k_y$.
As depicted in Fig. \ref{fig:band}. A,
the band minima are located at  $\vec K_{px}=(\frac{\pi}{a_0},0)$ for 
the $p_x$-band and $\vec K_{py}=(0,\frac{\pi}{a_0})$ for the $p_y$-band, 
respectively.

In the triangular lattice, we set the $\pi$-bonding to zero 
which does not change the qualitative band structures.
We take unit vectors from one site to its six neighbors
as $\pm \hat e_{1,2,3}$ with $\hat e_1=\hat e_x$,
$\hat e_{2,3}=-\frac{1}{2} \hat e_x \pm\frac{\sqrt{3}}{2} \hat e_y$.
The Brillouin zone takes the shape of a regular hexagon with the
edge length $4\pi/(3 a)$. 
The energy spectrum of $H_0$ is
\bea
E(k)= t_\pp \Big\{ f_{\vec k} \mp \sqrt{f^2_{\vec k}
-3g_{\vec k}} \Big\},
\eea
with
\bea
f_{\vec k}= \sum_{i=1}^3 \cos (\vec k \cdot \hat e_i), \ \ \
g_{\vec k} =\sum_{3\ge i>j\ge 1} \cos(\vec k \cdot \hat e_i) 
\cos (\vec k \cdot \hat e_j).
\eea
The spectrum contains three degenerate minima  located at
the three non-equivalent middle points of the edges as
$
K_1=(0, \frac{2\pi}{\sqrt 3 a}), 
K_{2,3}=(\pm\frac{\pi}{a}, \frac{\pi}{\sqrt 3 a})$,
The factor $e^{i \vec K_1 \cdot \vec r}$ takes the value of 
$\pm 1$ uniformly in each horizontal row but alternating
in adjacent rows.
If the above pattern is rotated at angles of $\pm\frac{2\pi}{3}$,
then we arrive at the patterns of $e^{i \vec K_{2,3} \cdot \vec r}$.
Each eigenvector is a 2-component superposition vector of $p_x$ and $p_y$
orbitals.
The eigenvectors at energy minima are
$\psi_{K_1}= e^{i\vec K_1 \cdot \vec r} |p_y\rangle$.
$\psi_{K_{2,3}}$ can be obtained by rotating 
$\psi_1$ at angles of $\pm \frac{2\pi}{3}$ respectively
as depicted in Fig. \ref{fig:band} B.

%---------------------------------------------------------------------
%----------------------------------------------------------------------
\subsection{Weak coupling analysis -- BEC at nonzero momenta}
If there were no interactions, any linear superposition of the
above band minima would be a valid condensate wavefunctions.
However, interactions will select a particular type combination
which exhibits TR symmetry breaking and agrees with
the above picture of ``orbital Hund's rule''.

\begin{figure}
\centering\epsfig{file=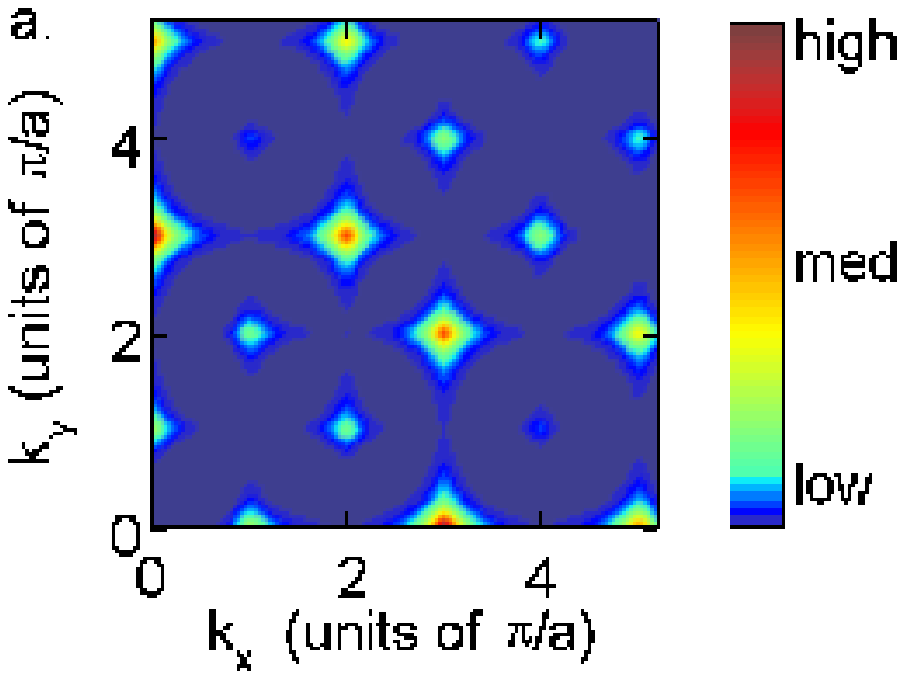,clip=1,width=0.40\linewidth,angle=0}
\hspace{5mm}
\centering\epsfig{file=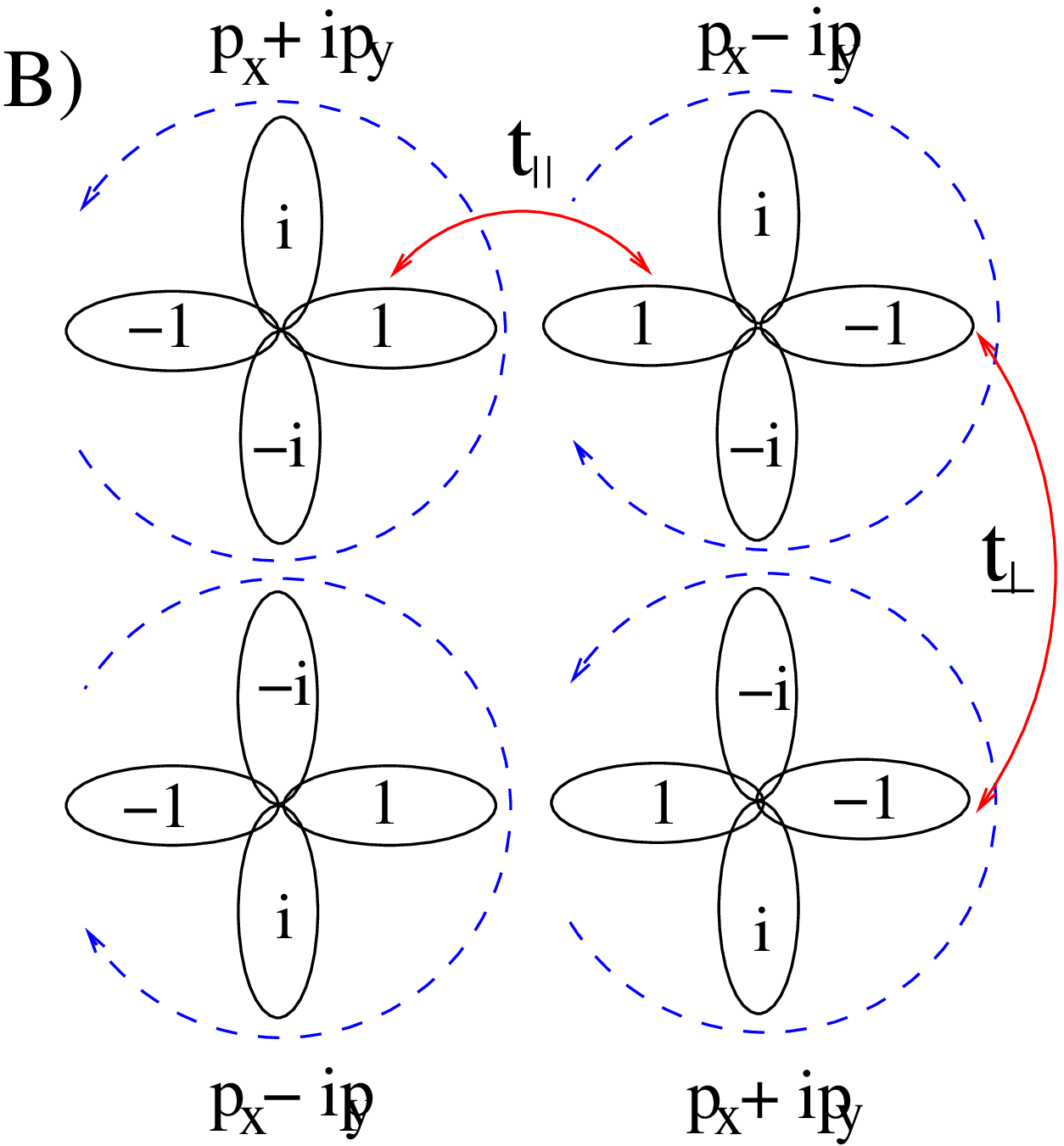,clip=1,width=0.30\linewidth,angle=0}
\caption{
A) The time-of-flight spectra of the $p$-orbital Bose
condensation in the square lattice.
The coherence peaks occurs at
$(m \frac{\pi}{a_0}, (n+\frac{1}{2})\frac{\pi}{a_0})$,
and $((m+\frac{1}{2})\frac{\pi}{a_0}, n \frac{\pi}{a_0})$. 
B) The staggered ordering of OAM moments in the square lattice
with the phase pattern on each site.
Each of the $\sigma$ and $\pi$-bonds achieves phase coherence.
From Liu and Wu, Ref 2. %\cite{liu2006}.
}
\label{fig:square}
\end{figure}

\subsubsection{The square lattice -- the staggered ordering of OAM moments}
In the square lattice, we take the condensate wavefunction
as $\Psi_{sq}(\vec r)= c_1 \psi_{p_x}(K_{p_x})+c_2 \psi_{p_y} (K_{p_y})$
under the constraint of $|c_1|^2+|c_2|^2=1$.
Any choice of $c_{1,2}$ minimizes the kinetic energy. 
However, the interaction $U$-terms break the degeneracy.
The degenerate perturbation theory shows that the ground state
values of $c_{1,2}$ take $c_1=1, c_2=i$ or its equivalent
TR partner of $c_1=1, c_2=-i$.
The mean field condensate can be described as
\bea
\frac{1}{\sqrt {N_0 !}}\Big\{\frac{1}{\sqrt 2} 
(\psi^\dagger_{K_{p_x}} +i \psi^\dagger_{K_{p_y}})\Big\}^{N_0}
|0\rangle,
\label{eq:square}
\eea
where $N_0$ is the particle number in the condensate.

For better insight, we transform the above momentum space condensate to 
the real space.
The orbital configuration on each site reads
\bea
e^{i\phi_{\vec r}}(|p_x\rangle +i\sigma_{\vec r} |p_y\rangle),
\eea
where the $U(1)$ phase $e^{i\phi_{\vec r}}$ is specified at the
right lobe of the $p$-orbital.
The Ising variable $\sigma_{\vec r}=\pm1$ denotes 
the direction of the OAM, and is represented by the anti-clockwise 
(clockwise) arrow on each site in Fig. \ref{fig:square} B.
Each site exhibits a nonzero OAM moment and breaks TR symmetry.
The condensate wavefunction of Eq. \ref{eq:square} describes the
staggered ordering of $\sigma_{\vec r}$.
We check that the phase difference is zero along each bond,
and thus no inter-site bond current exists.

The condensate described in Eq. \ref{eq:square} breaks both TR reversal and 
translational symmetries, and thus  corresponds to a BEC at non-zero momenta.
This feature should exhibit itself in the time-of-flight experiments
which have been widely used to probe the momentum distribution of cold atoms.
The coherence peaks of Eq. \ref{eq:square} are not located 
at integer values of the reciprocal lattice vectors but at
$(m \frac{\pi}{a_0}, (n+\frac{1}{2})\frac{\pi}{a_0})$,
and $((m+\frac{1}{2})\frac{\pi}{a_0}, n \frac{\pi}{a_0})$
as depicted in Fig. \ref{fig:square} A.
Furthermore unlike the $s$-orbital condensate the $p$-wave Wannier function
superposes a {\it non-trivial profile} on the height of density peaks. 
As a result, the highest peaks are shifted
from the origin---a standard for the $s$-wave peak---to the
reciprocal lattice vectors 
whose magnitude is around $1/l$ where $l$ is the characteristic
length scale of the harmonic potential of each optical site.
A detailed calculation of form factors is given in Ref. \cite{liu2006}.

%--------------------------------------------------------------------
\subsubsection{Excitations --  gapless phonons and gapped orbital modes}
The elementary excitations in $p$-orbital BECs consist of
both the gapless phonon mode and the gapped orbital mode.
The former corresponds to the Goldstone mode related to the
$U(1)$ symmetry breaking, and the latter corresponds to
the flipping of the direction of OAM moments. 
In the following, we will take the 2D square lattice as
an example.
 
We assume the condensate as $\frac{1}{\sqrt 2}[p^\dagger_x (Q_x) 
+ i p^\dagger_y (Q_y)]$ with $Q_x=(\pi, 0)$ and $Q_y=(0,\pi)$.
The boson operators take the expectation values as
\bea
\avg{|p_x(\vec r)|}= (-1)^{r_x} \phi, \ \ \
\avg{|p_y(\vec r)|}=i(-1)^{r_y} \phi.
\eea
From minimizing the onsite part of the free energy respect to the
condensate order parameter $\phi$
\bea
F=-\mu n+\frac{U}{2}( n^2-\frac{1}{3} L_z^2),
\eea
we have $\mu=\frac{4U}{3} |\phi|^2$ respect to the band minima.
The fluctuation around the expectation value is defined as
\bea
p_x(\vec r)=\avg{|p_x(\vec r)}+\delta p_x, \ \ \
p_y(\vec r)=\avg{|p_y(\vec r)}+\delta p_y.
\eea
Then the interaction terms in Eq. \ref{eq:bshamint} are expanded as 
\bea
H_{int}&=&\frac{4}{3} |\phi|^4
+[\delta p^\dagger_x \delta p_x +\delta p^\dagger_y \delta p_y]
\frac{8}{3} |\phi|^2
+ \frac{1}{3} \phi^{*,2} [\delta p_x \delta p_x
-\delta p_y \delta p_y ]  \nn \\
&+& \frac{1}{3} \phi^2  [ \delta p_x^\dagger \delta p_x^\dagger
-\delta p_y^\dagger \delta p_y^\dagger ]
+\frac{2i}{3} (-)^{r_x+r_y} 
[ \phi^2 \delta p_x^\dagger \delta p_y^\dagger
- \phi^{*,2} \delta p_x \delta p_y ].
\eea
Combined with the free part, we arrive the mean field
Hamiltonian as
\bea
H_{MF}=\sum_{\vec k, a,b}^\prime \Psi^\dagger (\vec k)_a M_{ab}(\vec k)
\Psi_b(\vec k)
\eea
where $\Psi^\dagger (\vec k)=
[ p^\dagger_x (\vec k), p^\dagger_y(\vec k+ \vec Q), p_x (-\vec k),
p_y (-\vec k - \vec Q) ]$, $\vec Q=(\pi, \pi)=\vec Q_x-\vec Q_y$,
and the summation is only over half of the Brillouin zone.
The matrix kernel reads
\bea
M(\vec k)=\left( \begin{array}{cccc}
\epsilon_x(\vec k)+\frac{4 U}{3}|\phi|^2 & 0 & \frac{2 U}{3} \phi^2 &
i\frac{2 U} {3} \phi^2\\
0 & \epsilon_y (\vec k +\vec Q)+\frac{4 U}{3} |\phi|^2 &
i\frac{2 U} {3} \phi^2 & \frac{2 U} {3} \phi^2 \\
\frac{2}{3} U \phi^{*,2} & -\frac{2 i}{3} U \phi^{*,2}&
\epsilon_x(\vec k)+\frac{4 U}{3}|\phi|^2&0 \\
 -\frac{2 i}{3} U \phi^{*,2} & \frac{2}{3} U \phi^{*,2}&
0& \epsilon_y (\vec k +\vec Q)+\frac{4 U}{3} |\phi|^2
\end{array} \nn
\right). 
\label{eq:kernel}
\eea
The spectra are reduced to generalized eigenvalue
problem of 
\bea
X^{-1} \big\{\mbox{diag}(1,1,-1,-1) ~M (\vec k) \big\} X =
 \mbox{diag}(E_1,E_2,-E_3,-E_4) ,
\eea
where $E_{1,2,3,4} (\vec k)$ are excitation eigenvalues,
and $X$ contains the eigenvectors.

Let us consider the excitation close to the condensation 
wavevector $\vec k=\vec Q_x+\vec q$, and then $\vec k+\vec Q
=\vec Q_y+\vec q$.
At small value of $\vec q$, we obtain the excitation spectra as
\bea
E_{1,3}(\vec k)&=&\sqrt{\bar \epsilon(\vec q) 
(\bar \epsilon (\vec q) +\frac{8}{3} U |\phi|^2 ) }, \ \ \ 
E_{2,4}(\vec k)= \bar \epsilon (\vec q) 
+\frac{4}{3} U |\phi|^2.
\eea
where $\bar \epsilon(\vec q)=\frac{1}{2}[\epsilon_x(\vec Q_x +\vec q)+
\epsilon_y(\vec Q_y+ \vec q)-\epsilon_x(\vec Q_x) -\epsilon_y (\vec Q_y)
] \approx \frac{t_\pp +t_\perp}{2}
(q_x^2 + q_y^2)$.
Clearly, the gapless mode with the linear dispersion relation
describes the superfluid phase fluctuations;
the gapped mode describes the orbital excitations corresponding
to the flipping of orbital angular momenta.

%-----------------------------------------------------------
\subsubsection{The triangular lattice -- the stripe ordering of OAM moments}

\begin{figure}
\centering\epsfig{file=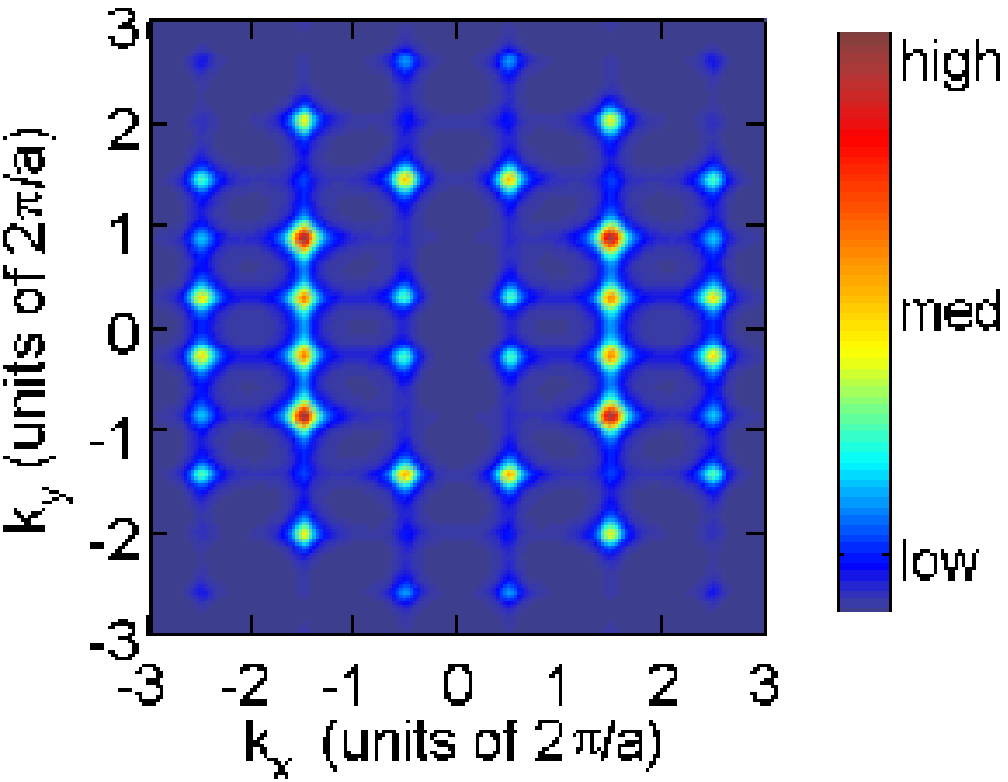,clip=1,width=0.35\linewidth,angle=0}
\hspace{3mm}
\centering\epsfig{file=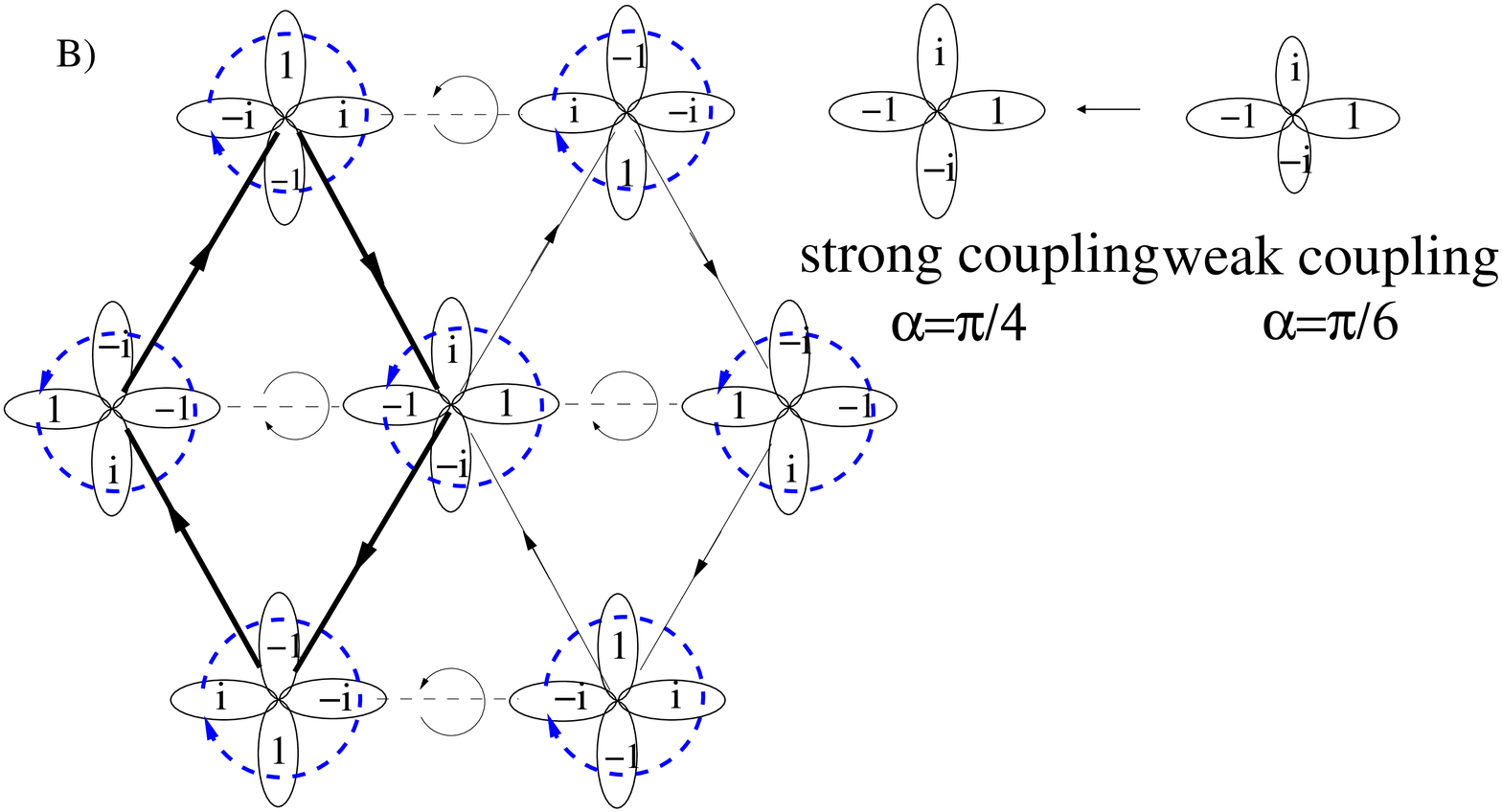,clip=1,width=0.51\linewidth,angle=0}
\caption{
A) The time-of-flight spectra of the $p$-orbital boson 
condensation in the triangular lattice.
The coherence peaks occurs at
B) The stripe ordering of OAM moments in the triangular lattice.
From Wu {\it et al.} Ref. 3. %\cite{wu2006}.
}
\label{fig:triangular}
\end{figure}

In the triangular lattice, the OAM moments instead form a stripe ordering,
i.e., the OAM moments along one row polarizes along the $z$-axis and
those along the neighouring rows polarizes with the opposite direction.
This can be intuitively understood as follows.
In the superfluid state, the OAM moments behave like vortices whose
interactions are long range.
The above stripe configuration of positive and negative vortices 
is the optimal configuration to minimize the globe vorticity.

Let us first examine the weak coupling limit.
Again we write a general form for condensation wavefunction 
as a linear superposition of the three  band minima
\bea
\Psi_c(\vec r)= \frac{1}{\sqrt N}
(c_1 \psi_{K_1} +c_2 \psi_{K_2} +c_3 \psi_{K_3}),
\eea
where $K_{1,2,3}$ are the locations of band minima defined in
Sect. \ref{subsect:band}.
Without loss of generality, we set $c_1=1$, 
and define $x=(c_2+c_3)/2$,  $y=(c_2-c_3)/2$ and $x=x_1+ix_2$ and $y=y_1+iy_2$,
then the normalization factor $N=\sqrt{1+2(x_1^2+y_1^2+x_2^2+y_2^2)}$.
The interaction energy persite is calculated as
\bea
E_{int}&=&\frac{1}{N^2} 
\big\{ (x_1^2+x_2^2-y_1^2-y_2^2)^2+8 x_1^2+8 y_1^2 \big\} +1\nn \\
&-& \frac{1}{N^2} \big\{ 4 (x_1 y_2 -x_2 y_1)^2 + 2x_2^2 + 2 y_2^2\big\}.
\eea
The terms in the first line are from the density-density interaction which
can be minimized by setting $x_1=y_1=0$ and $x_2=\pm y_2$.
This means only one of $c_{2,3}$ is nonzero and purely imaginary.
In this case, the particle number on each site is uniform.
The terms in the second line can also be minimized at this
condition with a further requirement of $x_2=\pm y_2=\pm \frac{1}{2}$.
If we take $c_2$ nonzero, then  $c_2=\pm i$.
Thus the mean field condensate can be expressed as
$
\frac{1}{\sqrt {N_0 !}}\Big\{\frac{1}{\sqrt 2} (\psi^\dagger_{K_2} +i \psi^\dagger_{K_3})\Big\}^{N_0}
|0\rangle
$
with $|0\rangle$ the  vacuum state and
$N_0$ the particle number in the condensate.
This state breaks the $U(1)$ gauge symmetry, as well as 
TR and lattice rotation symmetries, thus the
ground state manifold is $U(1)\otimes Z_2 \otimes Z_3$.
This state also breaks lattice translation symmetry, 
which is, however, equivalent to suitable combinations of 
$U(1)$ and lattice rotation operations.

Again we transform the above momentum space condensate to 
the real space, whose orbital configuration takes
\bea
e^{i\phi_{\vec r}}(\cos \alpha |p_x\rangle +i\sigma_{\vec r}
\sin \alpha |p_y\rangle)
\label{eq:triangular}
\eea
with $\alpha=\frac{\pi}{6}$ as $U/t\rightarrow 0$.
The general configuration of $\alpha$  is
depicted in Fig. \ref{fig:triangular} B for later convenience.
At $U/t\rightarrow 0$, $p_{x,y}$  are not equally populated, 
and the moment per particle is $\frac{\sqrt{3}}{2} \hbar$.
This does not fully optimize  $H_{int}$
which requires $L_{z,\vec r}=\pm \hbar$.
However, it fully optimizes $H_0$ which 
dominates over $H_{int}$ in the weak coupling limit.
We check that the phase difference is zero along each bond,
and thus no inter-site bond current exists.

Interestingly, as depicted in Fig. \ref{fig:triangular} B,  OAM moments
form a stripe order along each horizontal row.
This stripe ordering in the weak coupling limit is robust at
small values of the $\pi$-bonding $t_\perp$ because it 
does not change the location of the band minima and 
the corresponding eigenfunctions of $\Psi_{K_{1,2,3}}$ at all. 

The driving force for this stripe formation in the SF
regime is the kinetic energy, i.e., the phase coherence between 
bosons in each site.
By contrast, the stripe formation 
in high T$_c$ cuprates 
is driven by the competition between long range repulsion and the
short range attraction in the interaction terms
\cite{kivelson2003}.

This stripe phase should manifest itself in the time of flight 
(TOF) signal as depicted in Fig.  \ref{fig:triangular} A.
In the superfluid state, we assume the stripe ordering wavevector $K_1$,
and the corresponding condensation wavevectors at $K_{2,3}$.
As a result, the TOF density peak position after a fight time of $t$
is shifted from the reciprocal lattice vectors $\vec G$ as follows
\bea
\avg{n(\vec r)}_t &\propto& \sum_{\vec G}\Big\{ |\phi_2 (\alpha, \vec k)|^2 
\delta^2(\vec k-\vec K_2- \vec G)
+|\phi_3 (\alpha, \vec k)|^2 \delta^2(\vec k-\vec K_3-\vec G)\Big\},
\eea
where $\vec k=m \vec r /( \hbar t)$; $\phi_{2,3}(\alpha,\vec k)$ is
the Fourier transform of the Wannier $p$-orbital wavefunction 
$|\phi_{2,3}(\alpha)\rangle$,
and  $\vec G = \frac{2\pi}{a}[m, (-m+2n)/\sqrt 3]$ with $m,n$ integers.
Thus Bragg peaks should occur at $\frac{2\pi}{a}
[m\pm\frac{1}{2}, \frac{1}{\sqrt 3} (-m+2n+\frac{1}{2})]$.
Due to the form factors of the $p$-wave Wannier orbit wavefunction
$|\phi_{2,3}(\alpha, \vec k)|^2$, the locations of the highest peaks 
is not located at the origin but around $|k|\approx 1/l_{x,y}$.
Due to the breaking of lattice rotation symmetry,
the pattern of Bragg peaks can be rotated at angles of
$\pm \frac{2\pi}{3}$.

%--------------------------------------------------------------------
%-------------------------------------------------------------------
\subsection{Strong coupling analysis -- the lattice gauge theory 
formalism}

\begin{figure}
\centering\epsfig{file=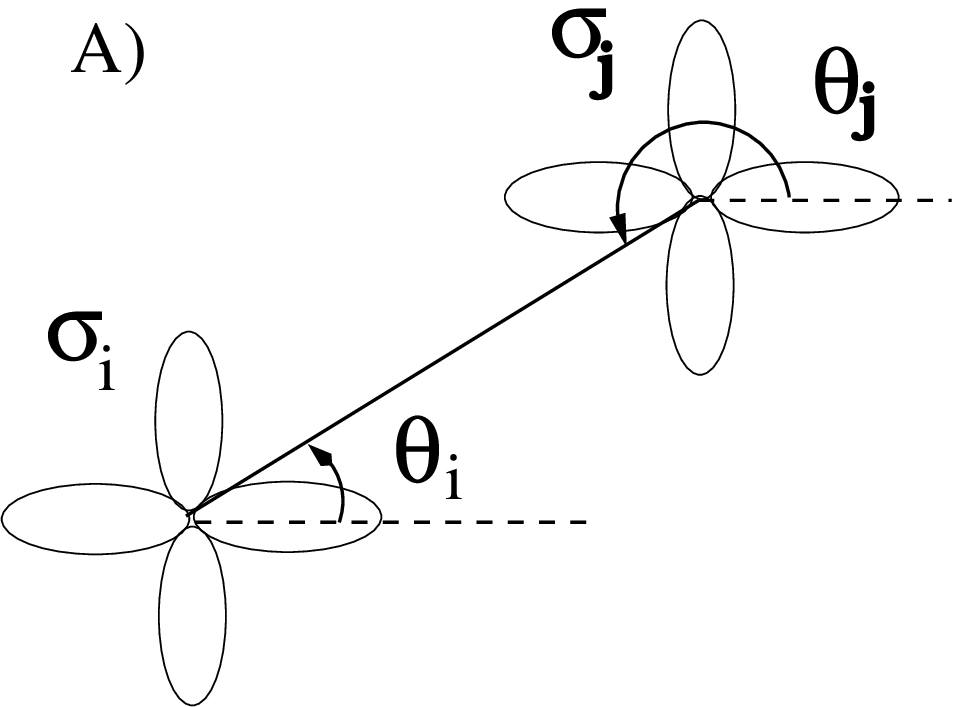,clip=1,width=0.37\linewidth,angle=0}
\hspace{5mm}
\centering\epsfig{file=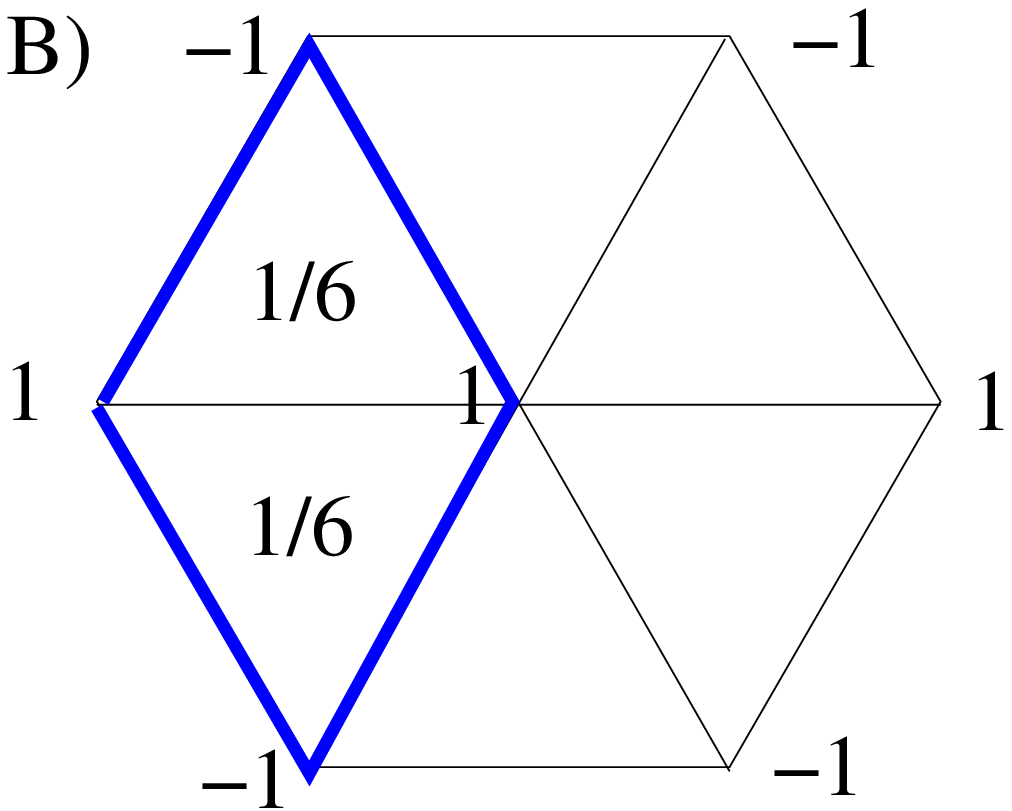,clip=1,width=0.34\linewidth,angle=0}
\caption{A) Strong coupling analysis for the inter-site coupling 
between OAM moments on neighboring sites.
The OAM moments are described by Ising variables $\sigma_{ij}$.
$\theta_{i}$ and $\theta_j$ are the azimuthal angles of the bonds
relative the $x$-axis.
B) The stripe ordering of OAM moments in the triangular lattice. 
The smallest unit is rhombic with the total vorticity of $1/3$. 
}
\label{fig:strong}
\end{figure}

In this subsection, the ordering of the OAM moments in the strong
coupling superfluid regime is examined.
We will employ the lattice gauge theory formalism developed by
Moore and Lee in Ref. \cite{moore2004} in the context of the
$p+ip$ Josephson junction array systems.

In the superfluid regime, each site $i$ is denoted by a $U(1)$ variable 
$\phi_i$ which denotes the superfluid phase along the $x$-direction,
and by an Ising variable $\sigma_{z,i}=\pm 1$ for the direction of the 
OAM moment as depicted in Fig. \ref{fig:strong}.
Along a general direction of the azimuthal angle $\theta$, the 
superfluid phase is $\phi_i+ \sigma_{z,i} \theta$.
The intersite $\sigma$ and $\pi$-bonding become the intersite 
Josephson coupling as
\bea
H_{ij}=-nt_\pp \cos(\phi_i-\phi_j-A_{\pp}(i,j)) 
      -nt_\perp \cos(\phi_i-\phi_j-A_{\perp} (i,j)),
\label{eq:josephson}
\eea
where $n$ is the average particle number per site.
The phase differences in Eq. \ref{eq:josephson} takes into account
the geometric orientation of the bond $\avg{ij}$
as captured by the gauge fields $A_\pp$ and $A_\perp$
with the definition that 
$A_{\pp}=\sigma_{z,i} \theta_i -\sigma_{z,j} \theta_j$,
$A_{\perp}=\sigma_{z, i} (\theta_i +\frac{\pi}{2})
-\sigma_{z,j} (\theta_j-\frac{\pi}{2})=A_{\pp}+\frac{\pi}{2} 
(\sigma_{z,i}+\sigma_{z,j})$,
and
$\theta_j=\theta_i+\pi$ are the azimuth angles relative to the $x$-axis
defined in Fig. \ref{fig:strong}.

Let us first consider the case of $t_\perp \neq 0$ and write down
an effective Hamiltonian for the Ising variables $\sigma_z$.
In order to  minimize both the Josephson couplings of the $\sigma$ 
and $\pi$-bonds, we need $\sigma_i=-\sigma_j$
and thus $A_{\pp, ij}=A_{\perp, ij}$.
Otherwise, if $\sigma_i=\sigma_j$, then $A_{\pp, ij}=A_{\perp, ij}+\pi$.
This costs an energy of $2t_\perp$, and leads to an effective 
antiferro-orbital interaction between the Ising variables as
\bea
H_{eff}=n t_\perp \sum_{ij} \sigma_{z,i} \sigma_{z,j}.
\eea
Thus the antiferro-orbital ordering of OAM moments
in the square lattice remains valid in the strong coupling limit
at non-vanishing $t_\perp$,
which enforces that both the $\sigma$ and $\pi$-bonds achieve phase coherence
as depicted in Fig. \ref{fig:square}. B.

On the other hand, in the limit of a vanishing $t_\perp$, the leading order 
effect involves multiple site interaction around a plaquette.
We follow the method described in Ref. \cite{moore2004},
and perform the duality transformation for the $U(1)$ phase
variables $\phi$ under the background of the geometric gauge 
potential $A_\pp$.
We separate the contributions from the phonon part and the vortex part as 
\bea
Z&=&Z_{ph} \sum_m^\prime \exp\Big\{-\frac{\pi^2 nt_\pp}{2}
\sum_x (m(x) - \Phi_x)^2\nonumber \\
&+&\pi nt_\pp \sum_{x\neq x^\prime} (m(x) -\Phi_x)
 \log \frac{|x-x^\prime|}{a}
(m (x^\prime) -\Phi_x^\prime)
\Big\},
\label{eq:vortex}
\eea
where $Z_{ph}$ is the phonon contribution in the vortex-free
configuration;
$x$ marks the dual lattice site (or the plaquette index in the original
lattice); $m$ is the vortex charge;
$\Phi_x$ is the external flux through the plaquette defined as
\bea
\Phi_x=\frac{1}{2\pi}\sum_{ij} A_{\pp, ij}.
\eea

In the square lattice, the geometric gauge flux around each plaquette
$x$ is calculated as
\bea
\Phi_x=-\frac{1}{4} (\sigma_i +\sigma_{i+e_x}+\sigma_{i+e_x+e_y}
+\sigma_{i+e_y}).
\eea
If $\Phi_x$ is integer-valued, it can be absorbed by shifting the zero
of the vortex charge as $m^\prime(x)=m(x) -\Phi_x$ which
remains integer-valued, thus there is no cost for
energy.
The feature can be captured by the effective Hamiltonian in Ref.
\cite{moore2004} as
\bea
H_{eff}=- K \sum_{ijkl} \sigma_i \sigma_i \sigma_k \sigma_l,
\label{eq:z2}
\eea
where $i,j,k,l$ are four sites around a plaquette centered at $x$;
$K\approx n t_\pp$ is the energy scale of the $\sigma$-bonding.
This model has a sub-extensive $Z_2$ symmetry investigated
in Ref. \cite{moore2004}, i.e., flipping the sign of $\sigma_z$
along each row or column leave Eq. \ref{eq:z2} invariant. 
Thus it cannot develop the ordering of OAM at any finite temperature.
Let us come back to the superfluid sector $\phi$, its most relevant
topological defect is the half-quantum vortices because 
$\Phi_x$ can take the values of $\pm \frac{1}{2}$.
The Kosterlitz-Thouless transition is associated with the unbinding
of half-quantum vortices.
As a result, the low temperature phase is the quasi-long range ordering
of pairing of bosons \cite{moore2004}.

Next we move to the strong coupling theory in the triangular lattice.
We start from the limit of $t_\perp=0$.
The geometrical gauge flux becomes
\bea
\Phi_x=-\frac{1}{6} (\sigma_1 +\sigma_{2}+\sigma_{3}),
\label{eq:triagflux}
\eea
where $1,2,3$ are three sites around a triangular plaquette.
There is no way to form an integer flux. 
The smallest vorticity per plaquette is $\pm \frac{1}{6}$ which
corresponds to either two $+1$'s and one $-1$, or two $-1$'s and one $+1$
to minimize the vortex core energy.
In such a dense vortex configuration, Eq. \ref{eq:vortex} rigorously speaking
does not apply because its validity relies on the assumption of small 
vortex fugacity. 
However, the structure of interactions among vortices
still implies that vortices form a regular lattice with alternating
positive and negative vorticity.
The dual lattice (the center of the triangular plaquette) is 
the bipartite honeycomb lattice.
It is tempting to assign $\pm \frac{1}{6}$ alternatively to each plaquette,
but actually it is not possible due to the following reason.
Consider a plaquette with vorticity $+\frac{1}{6}$, thus
its three vertices are  with two $+1$'s and one $-1$.
The neighboring plaquette sharing the edge with two $+1$'s must have the
same vorticity, and merges with the former one to form a rhombic plaquette
with vorticity $\frac{1}{3}$ as depicted in Fig. \ref{fig:strong} B. 
Thus the ground state should exhibit a staggered pattern of  rhombic
plaquette with vorticity of $\pm \frac{1}{3}$.
This arrangement precisely corresponds to the stripe order of
the Ising variables.
This configuration breaks both the lattice rotational and translational 
symmetries, which is six-fold degenerate.
If we turn on the small $t_\perp$ term, it results in an antiferromagnetic
Ising coupling with nearest bonds, the stripe configuration also
satisfies its ground state requirement.
Thus we believe that $t_\perp$ term does not change the stripe
configuration in the strong coupling limit either.

Having established the stripe order for the OAM moments, it is
straightforward to further optimize the $U(1)$ phase variable $\phi$.
The result is depicted in Fig. \ref{fig:triangular} B with $\phi$
marked on the right lobe of the $p_x\pm ip_y$ orbit.
Each horizontal bond has perfect phase match, while each of tilted bond
has a phase difference of $\frac{\pi}{12}$.
Thus around each rhombic plaquette, the phase winding is $\frac{\pi}{12} 
\times 4=\frac{\pi}{3}$ and gives rise to the Josephson supercurrent 
along the bonds as
\bea
j=\frac{t_\pp n}{2} \sin \Delta \theta, 
\label{eq:pqcurrent}
\eea
with $\Delta\theta=\frac{\pi}{6}$ and
the directions specified by arrows in Fig. \ref{fig:triangular} B.
In other words, in addition to the stripe order of the Ising variables
which corresponds to the onsite OAM moments, there exists a staggered
plaquette bond current order.
This feature does not appear in the square lattice \cite{liu2006}.
The orbital plaquette current ordering has been studied in the strongly
correlated electron systems, such as the circulating orbital current 
phase \cite{varma1997} and the d-density wave states in the high T$_c$ compounds
\cite{chakravarty2001}.
It is amazing that in spite of very different microscopic mechanism
and energy scales, the two completely different systems exhibit a 
similar phenomenology.

%---------------------------------------------------------------------
%---------------------------------------------------------------------

\subsection{Intermediate coupling regime in the triangular lattice -- 
self-consistent mean field analysis}

\begin{figure}
\centering\epsfig{file=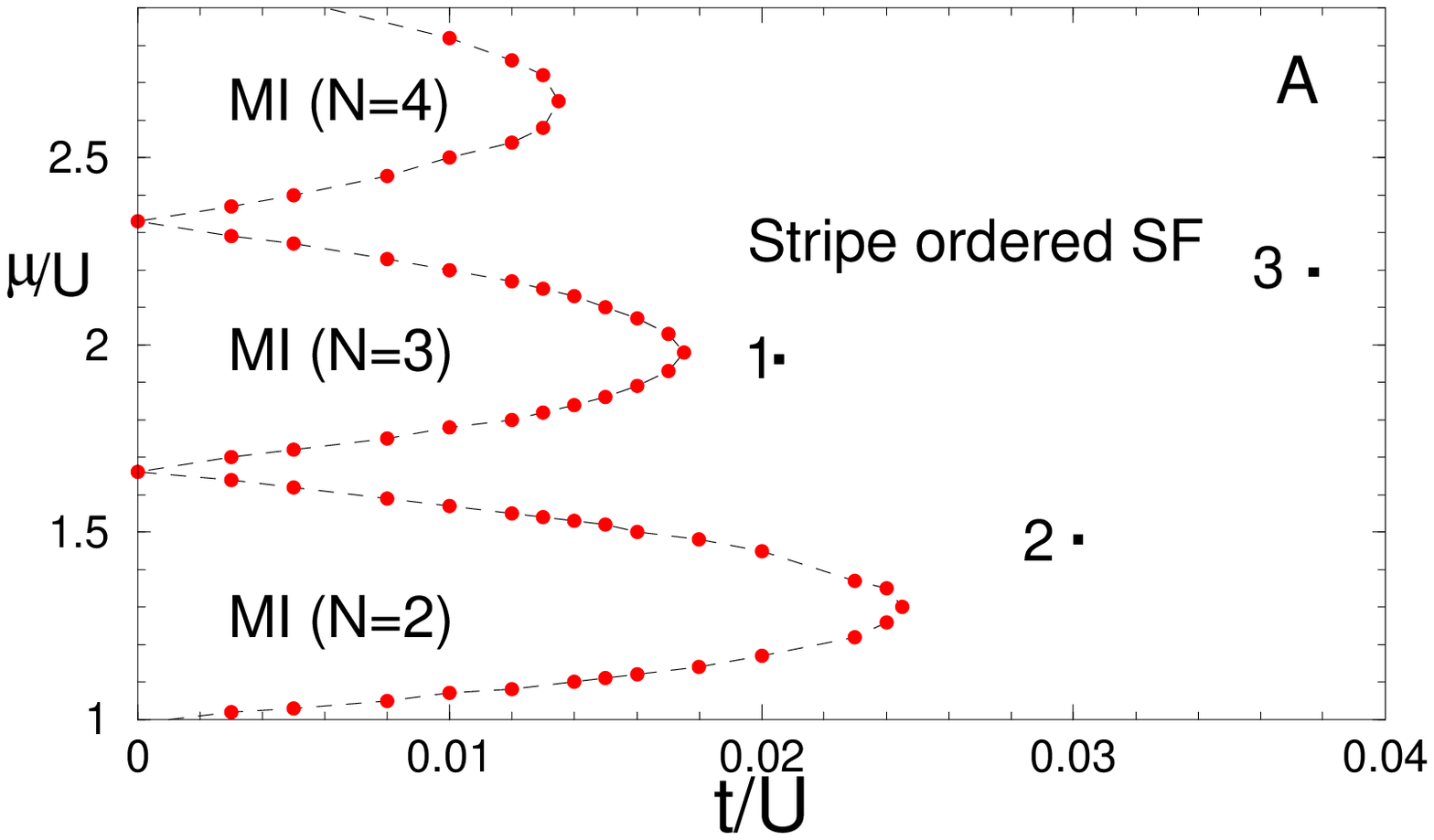,clip=1,width=0.48\linewidth,angle=0}
\centering\epsfig{file=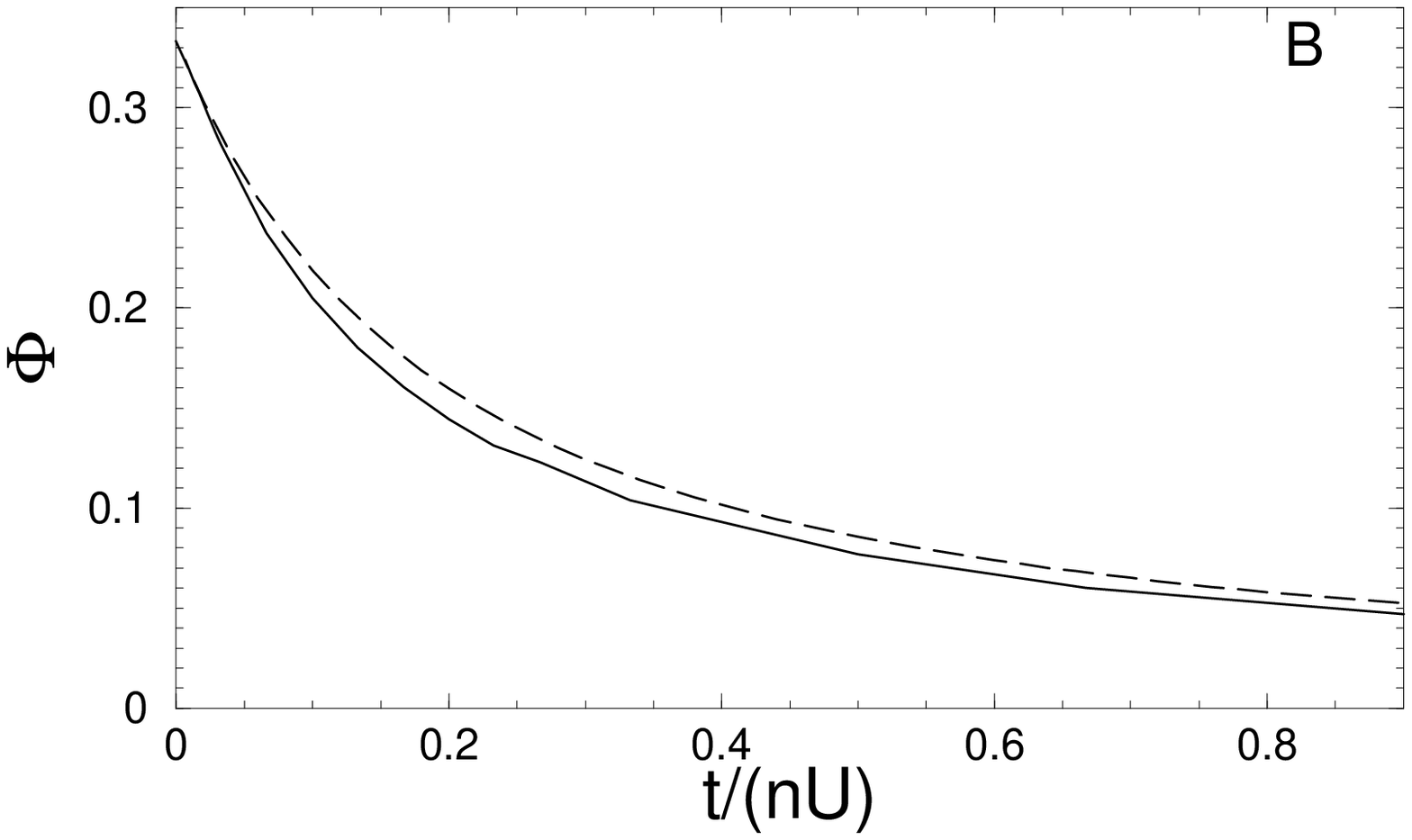,clip=1,width=0.48\linewidth,angle=0}
\caption{A) Phase diagram based on the GMF theory
in the $2\times2$ unit cell (see Fig. \ref{fig:triangular} B).
Large scale Gutzwiller mean field calculations in a $30\times 30$ lattice are
performed to confirm the stripe ordered superfluid (SF) phase
at points 1, 2 and 3 with $(t/U,\mu/U)=(0.02,2),
(0.03,1.5)$ and $(0.038,2.2)$, respectively.
B) The flux $\Phi$ around a rhombic plaquette v.s. $t/(nU)$.
It decays from $\frac{1}{3}$ in the strong coupling limit to $0$ 
in the non-interaction limit.
The solid line is the Gutzwiller result at $n=3$, while the dashed line
is based on the energy function Eq. \ref{eq:tricond} of the
trial condensate. From Wu {\it et. al}, Ref. 3. %\cite{wu2006}.
}
\label{fig:triang-phase}
\end{figure}

We have shown that in the triangular lattice the stripe ordering of 
the OAM moments exist in both weak and strong coupling superfluid states.
The orbital configurations are slightly different: the orbital mixing
angle $\alpha$ defined in Eq. \ref{eq:triangular} equals to $\frac{\pi}{6}$
in the weak coupling limit, while it equals to $\frac{\pi}{4}$
in the strong coupling limit.
This arises from the competition between the kinetic energy 
and the onsite Hund's rule.
Below we will see that as $U/t_\pp$ goes from small to
large, the stripe ordering remains the same with a smooth
evolution of $\alpha$ from $\frac{\pi}{6}$ to $\frac{\pi}{4}$
as depicted in Fig. \ref{fig:triangular} B.

We use a Gutzwiller type mean field theory for a $30\times 30$ lattice 
under the periodic boundary condition.
We explicitly checked for three sets of parameters $(t_\pp/U, \mu/U)$
marked as points of $1, 2$ and $3$ in Fig. \ref{fig:triang-phase} B.
The stripe-ordered ground state with a $2\times 2$ unit cell depicted
in Fig. \ref{fig:triangular} A is found stable against small random 
perturbations in all the three cases.
Then we further apply the Gutzwiller type theory assuming a $2\times 2$ 
unit cell and obtain the phase diagram depicted in Fig. 
\ref{fig:triang-phase} A which includes both
the stripe ordered superfluid phase and Mott-insulating phases. 

In order to gain a better understanding of the numerical results,
we write the trial condensate with the $p$-orbital configuration on 
each site as
\bea
 e^{i\phi_{\vec r}} (\cos\alpha |p_x\rangle + i \sigma_{\vec r}
\sin\alpha |p_y\rangle).
\eea
We have checked that the optimal pattern for the $U(1)$ phase $\phi_{\vec r}$
does not depend on the orbital mixing angle of $\alpha$, and it 
also remains the same for all the coupling strength.
The phase mismatch $\Delta \theta$ defined in Eq. \ref{eq:pqcurrent} 
on the tilted bonds for a general orbital mixing angle $\alpha$
can be calculated through simple algebra as
\bea
\Delta\theta=2\gamma-\pi/2, \ \ \ 
\mbox{with} ~~ \tan \gamma={\sqrt 3} \tan \alpha,
\eea
and  the corresponding Josephson current is
$j= n t \sin \Delta \theta$.
The value of $\alpha$ is determined by the minimization of the
energy per particle of the trial condensate as
\bea
{\cal E}(\alpha) = -t [1+ 2 \sin (2 \alpha +\frac{\pi}{6})]- 
\frac{n U}{6} \sin^2 2 \alpha+\frac{n U}{3},
\label{eq:tricond}
\eea
where the first term is the contribution from the kinetic energy
which requires phase coherence, and the second term is the
interaction contribution  reflecting the Hund's rule physics.
In the strong and weak coupling limits, the energy minimum 
is located at $\alpha=\frac{\pi}{4}$ and  $\frac{\pi}{6}$,
respectively.
The corresponding fluxes in each rhombic plaquette $\Phi= 4\Delta \theta/(2\pi)=
0$ and $\pm\frac{1}{3}$ respectively, which agree with the previous analyses.
In the intermediate coupling regime, we present both results of $\Phi$
at $n=3$ based on the Gutzwiller mean field theory and 
those of Eq. (\ref{eq:tricond}) in Fig.~\ref{fig:triang-phase}B.
They agree with each other very well, and confirm the validity of
the trial condensate wavefunction.
Moreover, in the momentum space, the trial condensate for a general 
$\alpha$  can be expressed as
$\frac{1}{\sqrt {N_0 !}}\Big\{\frac{1}{\sqrt 2} (\psi^{\prime\dagger}_{K_2} 
+i \psi^{\prime\dagger}_{K_3})\Big\}^{N_0}|0\rangle,
\label{eq:generalcond}$
where $\psi^\prime_{K_{2,3}}(\vec r)= 
e^{i\vec K_{2,3} \cdot \vec r} |\phi_{2,3}(\alpha)\rangle$
with $|\phi_{2,3}(\alpha)\rangle= -\cos \alpha |p_x\rangle\mp \sin \alpha
|p_y\rangle$ respectively.

%----------------------------------------------------------------
%----------------------------------------------------------------
\subsection{Orbital angular momentum ordering in the Mott-insulating 
states}
\label{subsect:mott}

So far, we have only discussed the orbital ordering in the superfluid
states. 
Since the OAM moment is a different degree of freedom from the
superfluid phase, we expect that its ordering can 
survive even inside the Mott-insulating phases.
In this subsection, we continue to study the exchange physics of
orbital bosons and the related ordering of OAM moments in the
absence of the superfluidity order.
For simplicity, we will use the triangular lattice as an example.

We consider the Mott-insulating phases with $n$ spinless bosons
per-site and two degenerate orbitals of $p_{x}$ and $p_y$.
We define the TR doublets of all the particles in the states of
$p_{x}\pm i p_{y}$ as the eigenstates of the Ising operator $\sigma_z$
with the eigenvalues of $\pm 1$.
The Ising part of the effective exchange Hamilton occurs at the 
level of the second order perturbation theory, while the Ising 
variable flipping process occurs at the $2n$-th order perturbation
theory.
We consider the large-$n$ case in which the system is deeply 
inside the Ising anisotropy class, and no orbital-flip process occurs
at the leading order.
In the following, we will study the physics of
the $2,3,4$-site exchange processes.

%-----------------------------------------------------------
\subsubsection{The two-site exchange}
Let us consider the virtual hopping processes in the Mott-insulating
states along the bond depicted in
Fig. \ref{fig:strong} A.
Both the $\sigma$-bonding $t_\pp$ and $\pi$-bonding $t_\perp$ are kept.
With the definition of 
$p^\dagger_\pm=\frac{1}{\sqrt 2} (p_x^\dagger \pm ip_y^\dagger),$
we can express the hopping as
\bea
H_t&=&-\frac{1}{2}\sum_{\sigma\sigma^\prime}
\Big\{ t_\pp
p^\dagger_\sigma(j) p_{\sigma^\prime} (i) 
e^{-i(\sigma \theta_j-\sigma^\prime \theta_i)} 
+t_\perp  p^\dagger_\sigma(j) p_{\sigma^\prime} (i) 
e^{-i(\sigma-\sigma^\prime) (\theta_i +\frac{\pi}{2})} 
+h.c. \Big\}  \nn \\
&=&
\frac{t_\pp-t_\perp}{2} \sum_\sigma \big\{p^\dagger_{j,\sigma}p_{i\sigma}+h.c.\big \}
+\frac{t_\pp+t_\perp}{2} \sum_\sigma \big \{p^\dagger_{j,\sigma} p_{i, -\sigma}
e^{-2 i \sigma \theta_i}+h.c. \big\},
\label{eq:motthop}
\eea
where the definition of angles $\theta_{i},\theta_{j}$ follows the convention
depicted in Fig. \ref{fig:strong} A.

We calculate the energy shifts in both the ferro-orbital 
configurations ($\sigma_i=\sigma_j$) and the antiferro-orbital 
configurations ( $\sigma_i=\bar\sigma_j$) within the second order
perturbation theory.
The energy difference between these two configurations is
\bea
\Delta E_{FO}&=& -\frac{2 n (n+1) [(t_\pp-t_\perp)/2]^2}{\Delta E_1} 
-\frac{2 n [(t_\pp+t_\perp)/2]^2}{\Delta E_2} \nn \\
\Delta E_{AFO}&=& -\frac{2 n (n+1) [(t_\pp+t_\perp)/2]^2}{\Delta E_1} 
-\frac{2 n [(t_\pp-t_\perp)/2]^2}{\Delta E_2}, 
\eea
where $\Delta E_1= \frac{2}{3} U$ 
and $\Delta E_2= \frac{2}{3} U (n+1)$.
The antiferro-orbital configuration has lower energy, which arises 
from the fact that the orbital-flip process has a larger amplitude 
than that of the orbital non-flip process described in Eq. \ref{eq:motthop}.
This is because that the $\sigma$-bonding and $\pi$-bonding 
amplitudes have a $\pi$-phase shift.
The Ising coupling $J_{AFO}$ reads
\bea
H= J_{AFO} \sum_{ij} \sigma_z (i) \sigma_z (j)
\label{eq:2ptcl}
\eea
where $J_{AFO} \approx\frac{2 n^2 t_\perp t_\pp }{3 U}$
with only the leading order contribution kept.
If $t_\perp$ is set to zero,  $J_{AFO}$ varnishes.
This is clear from the fact that we can flip the
sign of  the $p$-orbit component perpendicular to the bond direction.
This operation changes the value of $\sigma_z$, but has no effect on the
energy.
In this case, we need to further study the multi-site virtual hopping
processes.

\subsubsection{The three-site ring exchange}
The $\sigma$-bonding term by itself gives non-zero ring exchange terms
for the multi-site processes, thus we neglect the
contribution from the $\pi$-bonding part.
In the following, we only keep the leading order virtual process
proportional to $(nt_\pp)^3/U^2$.

We consider a triangular plaquette with three sites $(i,j,k)$
each of which is denoted by the particle number and the Ising
variables as $(n,\sigma_{z,i}), (n,\sigma_{z,j}), (n,\sigma_{z,k})$, respectively.
There are 12 different ring-hopping processes whose contribution
is at the order of $(nt_\pp)^3/U^2$.
We enumerate 4 of them explicitly as
\bea
&&(n,\sigma_{z,i}) (n,\sigma_{z,j}) (n,\sigma_{z,k}) \rightarrow
(n-1,\sigma_{z,i}) (n+1,\sigma_{z,j}) (n,\sigma_{z,k})\nn \\
&\rightarrow&
(n-1,\sigma_{z,i}) (n,\sigma_{z,j}) (n+1,\sigma_{z,k}) 
\rightarrow(n,\sigma_{z,i}) (n,\sigma_{z,j}) (n,\sigma_{z,k}) \nn \\
&&(n,\sigma_{z,i}) (n,\sigma_{z,j}) (n,\sigma_{z,k}) \rightarrow
(n+1,\sigma_{z,i}) (n-1,\sigma_{z,j}) (n,\sigma_{z,k}) \nn \\ 
&\rightarrow&
(n,\sigma_{z,i}) (n-1,\sigma_{z,j}) (n+1,\sigma_{z,k}) 
\rightarrow
(n,\sigma_{z,i}) (n,\sigma_{z,j}) (n,\sigma_{z,k}) \nn \\
&&(n,\sigma_{z,i}) (n,\sigma_{z,j}) (n,\sigma_{z,k}) \rightarrow
(n-1,\sigma_{z,i}) (n+1,\sigma_{z,j}) (n,\sigma_{z,k}) \nn \\ 
&\rightarrow&
(n,\sigma_{z,i}) (n+1,\sigma_{z,j}) (n-1,\sigma_{z,k}) 
\rightarrow
(n,\sigma_{z,i}) (n,\sigma_{z,j}) (n,\sigma_{z,k}) \nn \\
&&(n,\sigma_{z,i}) (n,\sigma_{z,j}) (n,\sigma_{z,k}) \rightarrow
(n+1,\sigma_{z,i}) (n-1,\sigma_{z,j}) (n,\sigma_{z,k}) \nn \\
&\rightarrow&
(n-1,\sigma_{z,i}) (n,\sigma_{z,j}) (n+1,\sigma_{z,k}) 
\rightarrow
(n,\sigma_{z,i}) (n,\sigma_{z,j}) (n,\sigma_{z,k}). 
\eea
The other $8$ processes can be obtained by a cyclic permutation $i\rightarrow
j\rightarrow k$.
After sum over all these processes, the corresponding energy shift reads
\bea
\Delta E (\sigma_i, \sigma_j, \sigma_k)\approx 6 (\frac{-t_\pp}{2})^3
(n (n+1))^{3/2} \frac{e^{-i 2\pi \Phi_3}+h.c.}
{(\Delta E)^2} 
&&= -J_3 \cos(2\pi \Phi_3), 
\label{eq:3ptcl}
\eea
where $J_3= \frac{3  (nt)^3}{2 (\Delta E)^2}$ and
$\Delta E= \frac{2U}{3}$; $\Phi_3$ is defined as 
before in the superfluid case.
Again for each plaquette, $\Phi_3$ takes the value of $\pm\frac{1}{6}$
which correspond to $\sigma_z$s taking the values of two $1$s and one $-1$ 
or two $-1$s and one $1$.
Eq. \ref{eq:3ptcl} plays a similar role to vortex core energy
in Eq. \ref{eq:vortex}.
Eq. \ref{eq:3ptcl} by itself has a huge ground state degeneracy 
because the requirement is the same as that of the antiferromagnetic
Ising model.
We need to add the four-site process to lift the degeneracy.

\subsubsection{The four-site ring exchange}
The four-site process mimics the interaction between two
adjacent vortices.
By the spirit of perturbation theory, the 4-particle
process is further suppressed by a factor $\frac{ nt_\pp}{U}$,
thus the interaction between vortices should be of short
range.
This makes sense in the Mott insulating state due to the
condensation of vortices and the resulting screening
effect.

Again we only consider the effect from the $\sigma$-bonding term,
and only keep the leading order contribution at the level of 
$(nt)^4/U^3$.
Similarly, we have 48 different virtual hopping processes.
For a plaquette with vertices $(ijkl)$, the energy shift is
\bea
&&\Delta E(\sigma_i,\sigma_j,\sigma_k,\sigma_l) \approx
-J_4 \cos(2\pi \Phi_4),
\label{eq:4ptcl}
\eea
where $J_4=\frac{5(nt)^4}{3 (\Delta E)^3}$ .
In the square lattice, $\Phi_4=(\sigma_1+\sigma_2+\sigma_3+\sigma_4)/4$.
If $t_\perp=0$, Eq. \ref{eq:4ptcl} is the leading
order contribution, thus the effective Hamiltonian looks the
same as that in the superfluid case but with a reduced coupling
constant.
If $t_\perp \neq 0$, then the two-site exchange of Eq.
\ref{eq:2ptcl} results in the long range antiferro-orbital
order.

In the triangular lattice, the $\Phi_4$ of the four-site plaquette $(ijkl)$
just equals the sum of the flux of the two triangular plaquettes.
Since each triangular plaquette takes the flux value of $\pm \frac{1}{6}$,
the rhombic four-site plaquette can only take the flux value 
of $0$, $\pm\frac{1}{3}$.
The stripe ordering pattern gives the maximum number of the 
zero-flux four-site plaquettes, and thus is selected as the 
ground state.
The stripe ordering appears at temperatures $T < J_4$
and disappears at $ J_3 > T > J_4$ in which the three-site
exchange process dominates the physics.

%*******************************************************************
%******************************************************************* 
%*******************************************************************

\section{Exotic condensation of bosons with spin-orbit coupling}
\label{sect:spinorbit}

In this section, we discuss another route for ``complex-condensation''
of bosons, i.e., spinful bosons with spin-orbit (SO) coupling.
The proof of Feynman's ``no-node'' theorem crucially relies on the fact
that the kinetic energy of bosons only contains even power of
momentum, thus SO coupling which linearly depends on momentum
invalidates Feynman's proof. 
In the following, we consider the simplest example of the two-component 
bosons with the Rashba-like SO coupling and investigate their 
condensate in the inhomogeneous harmonic traps.
A related work on the BEC of two-component bosons has also been
studied by Stanescu {\it et al.} in Ref. \cite {stanescu2008}.

%We find that the ground state condensate develops orbital angular momentum,
%which can be viewed as the formation of the half-quantum vortex.
%The spin density distribution exhibits a cylindrically symmetric spiral
%pattern with the pitch vector determined by the SO coupling strength.
%We will also discuss the realization of these states in exciton
%systems.

%-----------------------------------------------------------------
%----------------------------------------------------------------
%\subsection{Half-quantum vortex and the skyrmion-like spin texture 
%in the harmonic trap}

\begin{figure}
\centering\epsfig{file=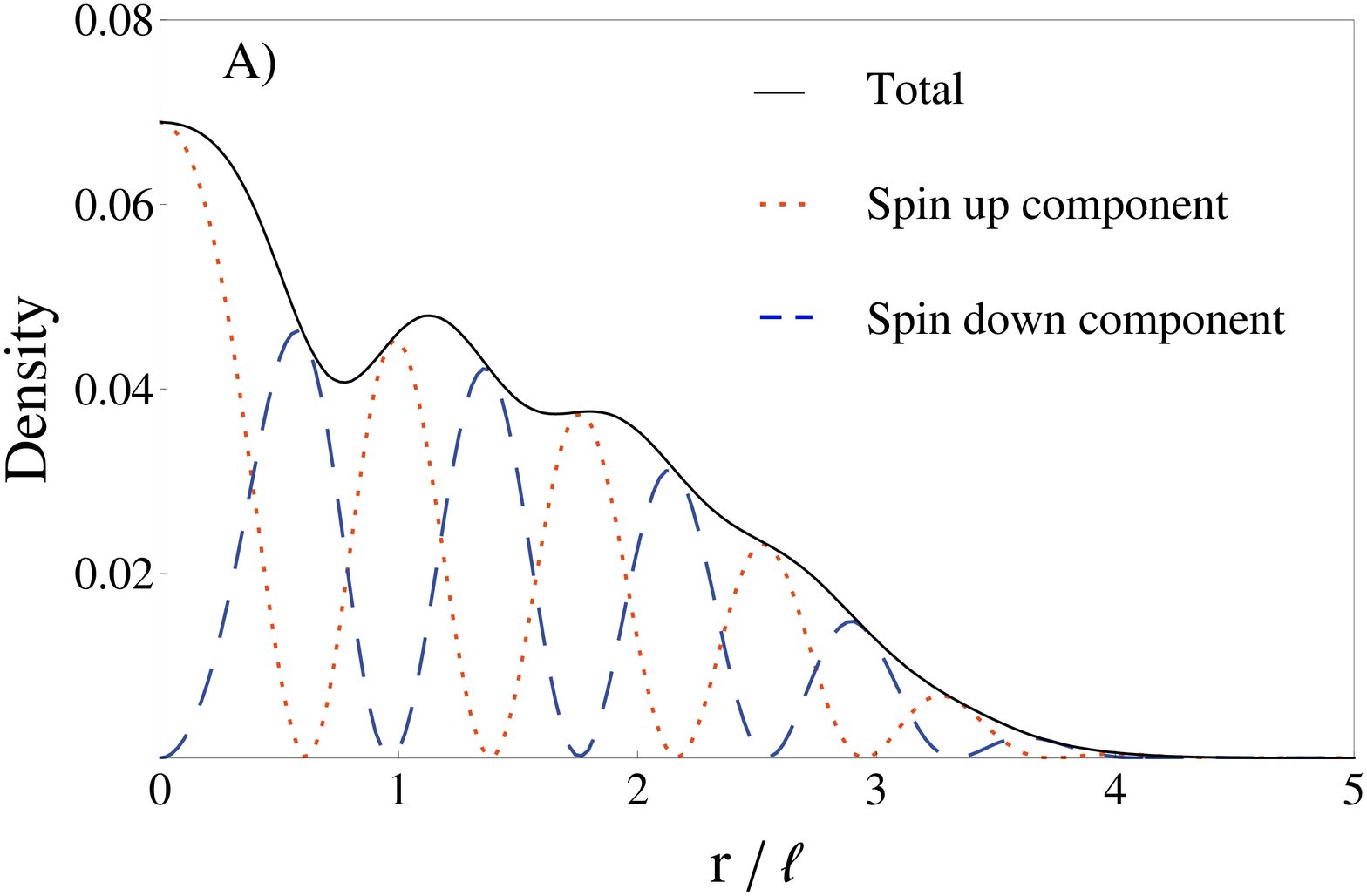,clip=1,width=0.5\linewidth,angle=0}
\centering\epsfig{file=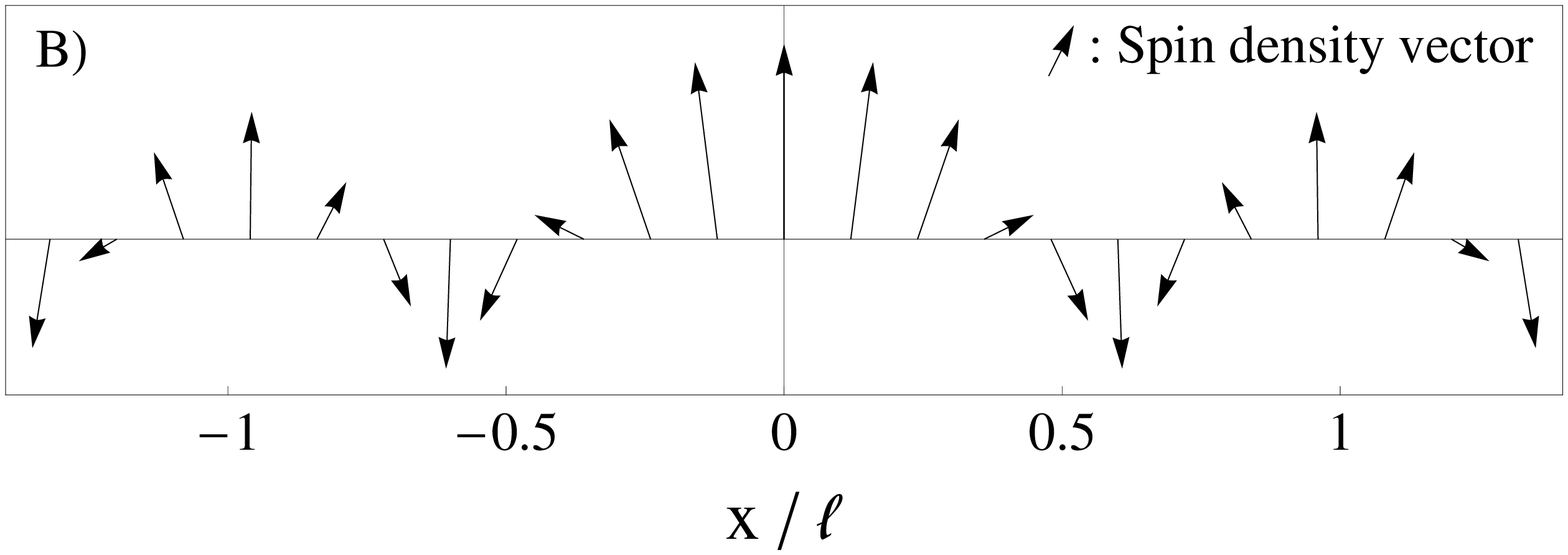,clip=1,width=0.5\linewidth,angle=0}
\caption{ A)The radial density distribution of spin up and down
components, and the total density distribution in the unit of $N_0$ 
at $\alpha=4$ and $\beta=40$. B) The spin density distribution 
along the $x$-axis which spirals in the $z$-$x$ plane at an 
approximate wavevector of $2k_0$, whose value at the origin
is normalized to 1. 
The spin density in the whole plane exhibits the skyrmion configuration.
From Wu {\it et al.} Ref. 5. %\cite{wu2008c}.
}
\label{fig:trap}
\end{figure}

We consider bosons with two-components denoted as pseudospin up and down.
The many-body Hamiltonian for interacting bosons is written as
\bea
H&=&\int d^2 \vec r~~ \psi^\dagger_\alpha
\big\{ -\frac{\hbar^2 \nabla^2}{2M}- \mu +V_{ext}(\vec r)\big\} \psi_\alpha
\nonumber \\
&+&\hbar\lambda_R  \psi^\dagger_\alpha (
-i \nabla_y \sigma_x  
+ i \nabla_x \sigma_y)\psi_\beta
+\frac{g}{2}\psi^\dagger_\alpha 
\psi^\dagger_\beta \psi_\beta \psi_\alpha, ~~~~~
\label{eq:mnbdy}
\eea
where $\psi_\alpha$ is the boson operator; $\alpha$
refers to boson pseudospin $\uparrow$ and $\downarrow$;
$V_{ext}$ is the external potential;
$g$ describes the $s$-wave scattering interaction which is assumed
to be spin-independent for simplicity; 
$\lambda_R$ is the Rashba SO coupling strength.
Although Eq. \ref{eq:mnbdy} is of bosons, it satisfies a suitably
defined TR symmetry like fermion systems as
$T=i\sigma_2 C$ with $T^2=-1$  where $C$ is the
complex conjugate and $\sigma_2$ operates on the pseudospin doublet. 
In the homogeneous system, the single particle
states are the helicity eigenstates of $\vec \sigma \cdot 
(\vec k \times \hat z)$ with the dispersion relations of
$\epsilon_{\pm} (\vec k)= 
\frac{\hbar^2}{2 M} (k \mp k_0)^2$ where $k_0=\frac{M \lambda_R}{\hbar}$.
The energy minima are located at the lower branch along a ring with 
radius $k_0$. 
The corresponding two-component wavefunction $\psi_{+}(\vec k)$ 
with $|k|=k_0$ can be solved as
$\psi_{+}^T (\vec k)= \frac{1}{\sqrt 2} (e^{-i \phi_k/2}, i e^{i \phi_k/2})$,
where $\phi_k$ is the azimuth angle of $\vec k$.

Instead of discussing the condensation in the homogeneous system in 
which frustrations occur due to the degeneracy of the single 
particle ground states,
we consider a 2D system with an external harmonic trap with 
$V_{ex}(r)=\frac{1}{2} M \omega_T^2 r^2$.
We define the characteristic SO energy scale for the
harmonic trap as $E_{so}= \hbar \lambda_R /l$ where 
$l=\sqrt{\hbar/(M\omega_T)}$ is the length scale of the trap, 
and correspondingly a dimensionless parameter as 
$\alpha=E_{so}/(\hbar \omega_T) =l k_0$. 
The characteristic interaction energy scale is defined as
$E_{int}=g N_0/(\pi l^2)$ where $N_0$ is the total particle number
and the dimensionless parameter $\beta=E_{int}/(\hbar \omega_T)$.

The Gross-Pitaevskii equation can be obtained as the saddle-point
equation of Eq. \ref{eq:mnbdy} as
\bea
\Big\{
-\frac{\hbar^2 \vec \nabla^2}{2M} &+&
\hbar \lambda_R (
-i \nabla_y \sigma_{x,\alpha\beta}  
+ i \nabla_x \sigma_{y,\alpha\beta})
+g (\psi^*_\gamma \psi_\gamma) 
+\frac{1}{2}M \omega^2_T r^2\Big\} \psi_\beta(r,\phi)\nn \\
&=&E \psi_\alpha(r,\phi).
\label{eq:GP}
\eea
Due to the 2D rotational symmetry, the ground state condensate wavefunctions 
can be denoted by the total angular momentum $j_z=L_z+\frac{1}{2} \sigma_z
=\pm\frac{1}{2}$, which can be represented in the polar coordinate as
\bea
|\psi_{\frac{1}{2}}\rangle=\left (\begin{array}{c}
f(r) \\
g(r) e^{i\phi}
\end{array}
\right ), \ \ \
|\psi_{-\frac{1}{2}}\rangle=\left (\begin{array}{c}
-g(r) e^{i\phi} \\ f(r)
\end{array}
\right),
\eea
where $r$ and $\phi$ are the radius and the azimuthal angle,
respectively. Both $f(r)$ and $g(r)$ are real functions.
Eq. \ref{eq:GP} is numerically solved
in the harmonic trap with parameters of $\alpha=4$ and $\beta=40$.
We choose the condensate as one of the TR doublet $|\psi_{\frac{1}{2}}\rangle$, 
and present its radial density profiles of both spin components $|f(r)|^2$ 
and $|g(r)|^2$ at $\alpha=4$ in Fig. \ref{fig:trap} A.
Further in this strong SO coupling case with $\alpha \gg 1$, 
the condensate wavefunction has nearly equal weight in the spin 
up and down components, i.e. $\int dr d\phi~ r|f(r)|^2 \approx 
\int dr d\phi~ r |g(r)|^2$,  thus the average spin moment along
the $z$-axis equals to zero. 
The total angular momentum per particle $j_z=\frac{\hbar}{2}$ is
mainly from the orbital angular momentum polarization, i.e.,
one spin component stays in the $s$-state and the other one in the 
$p_x+i p_y$-state.
This is an example of half-quantum vortex configuration 
\cite{salomaa1985,wu2005,zhou2003}, thus 
spontanously breaking TR symmetry.
Clearly this is a ``complex-valued'' ground state wavefunction 
beyond the ``no-node'' theorem.

This condensate wavefunction exhibits interesting spin density
distributions in real space as skyrmion-like spin textures.
The radial wavefunction in pseudo-spin up and down components
$f(r)$ and $g(r)$ exhibit oscillations with 
an approximate wavevector of $ k_0$,
which originates from the ring structure of the low energy states
in momentum space, thus are analogous to Friedel 
oscillations in fermion systems.
The pseudo-spin up component is $s$-wave like, thus $f(r)$
reaches the maximum at $r=0$; while the down component
is of the $p$-wave, thus $g(r)=0$ at $r=0$.
In other words, approximately there is a relative phase shift of 
$\frac{\pi}{2}$ between the oscillations of $f(r)$ and $g(r)$.
In Fig. \ref{fig:trap} A, $|f(r)|^2$ and $|g(r)|^2$ are plotted.
The spin density distribution $\vec S(r,\phi)=
\psi_{\frac{1}{2},\alpha}^* (r,\phi) \vec 
\sigma_{\alpha\beta} \psi_{\frac{1}{2},\beta}(r,\phi)$ can be expanded as
\bea
S_z (r,\phi) &=&\frac{1}{2} (|f(r)|^2 -|g(r)|^2 ), \ \ \
S_x (r, \phi)=f(r) g(r) \cos \phi, \nn \\
S_y (r, \phi)&=&  f(r) g(r) \sin \phi.
\eea
Along the $x$-axis, the spin density lies in the $z$-$x$ plane 
as depicted in Fig. \ref{fig:trap} B.
Because the circulating supercurrent is along the tangential direction,
the spin density distribution is along the radial direction and
exhibits an interesting topological texture configuration which
spirals in the $z$-$x$ plane at the pitch value of the density oscillations.
The distribution in the whole space can be obtained through a 
rotation around the $z$-axis.
This spin texture configuration is of the skyrmion-like.

%------------------------------------------------------------------
%------------------------------------------------------------------
%\subsection{Experimental realization}

Next we discuss the possible realization of the above exotic 
BEC in exciton systems.
Excitons are composite objects between conduction electrons and 
valence holes \cite{keldysh1986,nozieres1982,comte1982,snoke1990,butov2004a,butov2007,timofeev2007}.
In particular, the recently progress on the indirect exciton
systems greatly enhances the lift-time 
\cite{butov1994,butov1998,butov2001,butov2002}, which provides
a wonderful opportunity to investigate the exotic state of matter 
of the exciton condensation \cite{butov2004a}. 
The ordinary bosons are too heavy to exhibit the relativistic SO 
coupling in their center-of-motion.
Due to the small effective mass of excitons,  SO coupling 
can result in important consequences, including anisotropic 
electron-hole pairing \cite{hakioglu2007,can2008}, 
spin Hall effect of the center-of-mass motion of excitons  
\cite{wang2007,wang2008}, and the Berry phase effect on exciton 
condensation \cite{yao2008}.

We will show that the Rashba SO coupling in the electron band
also survives in the center-of-mass motion of excitions.
We begin with the Hamiltonian of indirect excitons
\bea
H_{e}&=&-\frac{\hbar^2}{2m^*_e} (\partial_{e,x}^2 +\partial_{e,y}^2)  +  
i \hbar \lambda_{R,el} (\partial_{e,x} \sigma_y - \partial_{e,y} \sigma_x),
\label{eq:hamrashba} \nonumber  \\
H_{hh}&=&-\frac{\hbar^2}{2m^*_{hh}} (\partial_{h,x}^2 +\partial_{h,y}^2),
\label{eq:hamhh} \nonumber \\
H_{e-hh}&=& -\frac{e^2}{\varepsilon \sqrt{|\vec r_e-\vec r_{hh}|^2 + d^2}},
\eea
where $m^*_{e}$ is the effective mass of conduction electrons;
$\lambda_{R,el}$ is the Rashba SO coupling strength of the conduction electron,
$\varepsilon$ is the dielectric constant; $d$ is the thickness of the
barrier.

For small exciton concentrations, we only need to consider
the heavy hole ($hh$) band with the effective 
mass $m_{hh}$ and $j_z=\pm\frac{3}{2}\hbar$, which is separated 
from the light hole band with a gap of the order of 10 meV.
We consider the center-of-mass motion in the BEC limit of excitons,
which can be separated from the relative motion in $H_{e}$ and $H_{hh}$.
Similarly to Ref. \cite{maialle1993}, the effective Hamiltonian 
of the 4-component $hh$ excitons denoted as 
$(s_e,j_{hh})=(\pm\frac{1}{2},\frac{3}{2}),(\pm\frac{1}{2},-\frac{3}{2})$ 
can be represented by the matrix form as
\bea
{\small H_{ex} = \left( \begin{array}{cccc}
E_{ex} (\vec k) & H_{so}(\vec k) & 0 &0 \\
H_{so}^* (\vec k) & E_{ex}(\vec k) +\Delta(\vec k)
&W(\vec k) &0 \\
0& W^*(\vec k) & E_{ex}(\vec k) +\Delta(\vec k) &
H_{so}(\vec k) \\
0& 0& H_{so}^*(\vec k) & E_{ex} (\vec k)
\end{array} 
\right ), \nonumber
}
\label{eq:hamexciton}
\eea
where $\vec k$ is the center-of-mass momentum;
$M=m_e^*+m_{hh}^*$ is the total mass of the exciton;
$E_{ex}(\vec k)=\hbar^2 k^2/(2M)$;
$H_{so}(\vec k)=-\frac{m_e^*}{M} \lambda_{R,el} (k_y+ i k_x)$;
$\Delta(\vec k)$ is the exchange integral 
and $W(\vec k)$ has the $d$-wave structure as $(k_x+i k_y)^2$,
both of which are exponentially suppressed by the tunneling
barrier for indirect excitons, and will be neglected below.
Consequentially, $H_{ex}$ becomes block-diagonalized.
We consider using circularly polarized light to pump the exciton of
$(-\frac{1}{2},\frac{3}{2})$, and then focus on the left-up $2 \times 2$
block 
of the $H_{ex}$ matrix with heavy hole spin $j_z=\frac{3}{2}$.
If we use the electron spin number $\uparrow$ and $\downarrow$
as the exciton component, we will arrive the Eq. \ref{eq:mnbdy}
with the renormalized SO coupling strength $\lambda_R=\lambda_{R,el}
m^*_e/M$.

We next justify the above choice of the values of parameters based on 
experimental situations.
The effective masses of electrons and holes in GaAs/AlGaAs quantum wells 
are $m_e^*\approx 0.07 m_e$ and $m_h^* \approx 0.18 m_e$ in
Ref. \cite{butov2000,butov2004a}.
The Rashba SO coupling strength  $\hbar \lambda_R$ can reach $1.8\times
10^{-12}$ eV.m in Ref. \cite{sih2005}.
Thus we can estimate a reasonable value of $k_0
=\frac{m_e^*\lambda_R}{\hbar} \approx 1.6\times 10^4$ cm$^{-1}$.
For a harmonic trap with $l=2.5\mu m$, $\alpha=k_0 l \approx 4$
and $\hbar\omega_T=\hbar^2/(M l^2)=0.5$ mK.
In two dimensional harmonic traps, the critical condensation
temperature $T_c\approx \hbar\omega_T\sqrt{N_0}$ \cite{dalfovo1999}.
If we take the exciton density $\rho=5\times 10^{10}$ cm$^{-2}$
and the effective area $\pi l^2$, we arrive at $T_c\approx 50$ mK, 
which is an experimentally available temperature scale \cite{butov2004}.
The average interaction energy per exciton in the typical density regime 
of $10^{10}$cm$^{-1}$ is estimated around 2meV in Ref. \cite{butov1999},
thus we take the interaction parameter $\beta=40$ in the calculation above.
The spatial periodicity of the spin texture is about $\pi/k_0\approx
2\mu m $ and, thus, is detectable by using optical methods.

\section{Conclusion}
\label{sect:conclusion}
We have reviewed the exotic condensations of bosons whose many-body 
wavefunctions are complex-valued in the coordinate representation, 
thus they go beyond the well-known paradigm of the ``{\it no-node}'' theorem.
We studied two possible ways to bypass this theorem to achieve
unconventional condensations, i.e., meta-stable
states of bosons in the high orbital bands in optical lattices,
and spinful bosons with SO coupling.

The first mechanism of orbital bosons is essentially an interaction 
effect which is characterized by the Hund's rule.
With the orbital degeneracy, bosons favor to enlarge their spatial
extension to reduce the inter-particle repulsion, which results in 
the maximization of their onsite OAM.
We reviewed the ordering of the OAM moments in both the square 
and triangular lattices due to the inter-site coupling in both
the weak and strong coupling limits.
The low energy excitations include both the gapless phonon mode
and the gapped orbital-flip mode.
The survival of the OAM ordering in the soft Mott-insulating 
regime is also discussed.
The second mechanism employing SO coupling, which linearly depends
on momentum, is a kinetic energy effect.
In this case, taking the absolute value of a non positive-definite 
wavefunction will change its energy and thus invalidate Feynman's proof.
We have shown that the condensate wavefunction in a harmonic trap 
becomes a half-quantum vortex and develops skyrmion-like spin textures.
In both cases, TR symmetry is spontaneously broken which are not
possible in the conventional BEC.

There are a number of interesting open issues for future study.
For example, the ordering of OAM moments in various different lattices
and in three dimensions 
generates a variety of challenging problems of the lattice
gauge theory \cite{wu2008d}.
More importantly, in order to facilitate the communication between
the theory work with cold atom experiments, detailed calculation and 
optimization of the life time of orbital bosons in different 
lattices are highly desired.
Furthermore, the dynamics of bosons in high orbital bands in another
challenging topic. 
Taking into account the tremendous progress of cold atom physics,
it would be great if this research can stimulate general interests on 
the unconventional condensates beyond the ``no-node'' theorem.

% Previously published material must be accompanied by written 
% permission from the author and publisher.

\section*{Acknowledgments}
I thank L. Balents, D. Bergman, H. H. Hung, W. C. Lee, J. Moore, 
I. Mordragon-Shem, W. V. Liu, S. Das Sarma, V. Stajonovic, and S. Z. Zhang
for fruitful collaborations on this topic.
In particular, I am grateful to W. V. Liu for his introducing me to 
this research direction, and S. Das Sarma for his guidance in the
research.
I  also thank I. Bloch, L. Butov, H. Deng, L. M. Duan, M. Fogler, J. Hirsch,
T. L. Ho, N. Kim, Z. Nussinov, L. Sham, Y. Yamamoto,
S. C. Zhang and F. Zhou for helpful discussions.
This work is supported by the NSF Grant No. DMR-0804775 and Sloan
Research Foundation.

%\bibliographystyle{prsty}
%\bibliography{orbital,exciton,spin32}

\end{document}